\documentclass[twocolumn,superscriptaddress,showpacs,nofootinbib,notitlepage,trackchanges]{aastex701}

\usepackage{graphicx,amsmath,amsfonts,amssymb,multirow,mathrsfs,ulem}
\usepackage{hyperref}
\usepackage{epstopdf}

\begin{document}

\title{Probing Baryonic Feedback Effect with CSST Weak Lensing and Future FRB Measurements}

\correspondingauthor{Yan Gong}

\email{Email: gongyan@bao.ac.cn}

\author[0009-0006-0803-0505]{Shuai Feng}
\affiliation{National Astronomical Observatories, Chinese Academy of Sciences, Beijing 100101, People's Republic of China}
\affiliation{University of Chinese Academy of Sciences, Beijing 100049, People's Republic of China}
\email{fshuai@bao.ac.cn}

\author[0000-0003-0709-0101]{Yan Gong}
\affiliation{National Astronomical Observatories, Chinese Academy of Sciences, Beijing 100101, People's Republic of China}
\affiliation{University of Chinese Academy of Sciences, Beijing 100049, People's Republic of China}
\affiliation{Science Center for CSST, National Astronomical Observatories, CAS, 20A Datun Road, Beijing 100101, China}
\email{gongyan@bao.ac.cn}

\author[0009-0009-1369-2476]{Qi Xiong}
\affiliation{National Astronomical Observatories, Chinese Academy of Sciences, Beijing 100101, People's Republic of China}
\affiliation{University of Chinese Academy of Sciences, Beijing 100049, People's Republic of China}
\email{xiongqi@bao.ac.cn}

\author[0000-0002-2552-7277]{Xiaohui Liu}
\affiliation{National Astronomical Observatories, Chinese Academy of Sciences, Beijing 100101, People's Republic of China}
\affiliation{University of Chinese Academy of Sciences, Beijing 100049, People's Republic of China}
\email{liuxh@bao.ac.cn}

\author[0000-0003-4528-656X]{Hengjie Lin}
\affiliation{National Astronomical Observatories, Chinese Academy of Sciences, Beijing 100101, People's Republic of China}
\affiliation{University of Chinese Academy of Sciences, Beijing 100049, People's Republic of China}
\email{liuxh@bao.ac.cn}

\author[0009-0009-2259-5221]{Dingao Hu}
\affiliation{National Astronomical Observatories, Chinese Academy of Sciences, Beijing 100101, People's Republic of China}
\affiliation{University of Chinese Academy of Sciences, Beijing 100049, People's Republic of China}
\email{liuxh@bao.ac.cn}

\author[0000-0001-6475-8863]{Xuelei Chen}
\affiliation{National Astronomical Observatories, Chinese Academy of Sciences, Beijing 100101, People's Republic of China}
\affiliation{University of Chinese Academy of Sciences, Beijing 100049, People's Republic of China}
\affiliation{Department of Physics, College of Sciences, Northeastern University, Shenyang 110819, China}
\affiliation{Centre for High Energy Physics, Peking University, Beijing 100871, People's Republic of China}
\affiliation{State Key Laboratory of Radio Astronomy and Technology, China}
\email{xuelei@nao.cas.cn}

\begin{abstract}
We explore the joint probe on the baryonic feedback effect using the weak lensing measurement from the upcoming Chinese Space-station Survey Telescope (CSST) photometric survey and the dispersion measure (DM) statistics of the  fast radio bursts (FRBs) from next-generation radio telescopes, i.e., the Square Kilometre Array (SKA) and the Deep Synoptic Array (DSA-2000). By employing the baryonic halo model, we compute the matter, electron, and matter--electron power spectra, and generate mock data considering realistic noise and systematic effects based on the designs of the telescopes. These mock data are then analyzed using the Markov Chain Monte Carlo (MCMC) method to investigate the parameter constraints. We find that CSST weak lensing alone can constrain the baryonic feedback parameter $\log_{10} T_{\text{AGN}}$ to an accuracy of $3.1\%$, with the sum of neutrino mass bound $\sum m_{\nu} < 0.53\,\mathrm{eV}$. When performing the $3\times2 \,\mathrm{pt}(e,\mathrm{m})$ analysis, the inclusion of FRB DM measurements can significantly improve the precision of $\log_{10} T_{\text{AGN}}$ to $0.4\%$, and will lead to a better constraint on $\sum m_{\nu}$ with an upper limit $< 0.47\,\mathrm{eV}$ by effectively breaking the degeneracy. Our results demonstrate that the joint observation of future FRB DM and weak lensing surveys is a powerful tool for probing the baryonic feedback effect, which is helpful in obtaining robust constraints on the neutrino mass and other important cosmological parameters.
\end{abstract}


\section{Introduction}
Weak lensing is one of the most powerful probes of the large-scale structure (LSS) of the Universe in cosmology. The gravitational potential of massive foreground structures induces subtle distortions in the shapes of background galaxies, which can be detected statistically over large samples of galaxies \citep{1992ApJ...388..272K,2000A&A...358...30V}. By analyzing the correlated distortions, weak lensing can precisely probe the nature of dark matter and dark energy, and enables stringent constraints on a wide range of cosmological models.
Current weak lensing surveys have already achieved stringent constraints on cosmological parameters, such as the Dark Energy Survey (DES; \citealp{2026arXiv260210065D}), the Kilo-Degree Survey (KiDS; \citealp{2025A&A...703A.158W}), and the Hyper Suprime-Cam (HSC; \citealp{2023PhRvD.108l3517M}) survey. With the operation of upcoming Stage IV surveys, e.g. the Chinese Space-station Survey Telescope (CSST) \citep{2026SCPMA..6939501C}, Vera Rubin Observatory's Legacy Survey of Space and Time (LSST) \citep{2019ApJ...873..111I}, and {\it Euclid} \citep{2025A&A...697A...1E}, the fundamental parameters governing evolution of the Universe are expected to be measured with unprecedented precision.

However, complex astrophysical processes related to baryons, such as gas cooling, star formation, supernova explosions, and feedback from active galactic nuclei (AGN), can significantly alter the distribution of matter at small scales \citep{2011MNRAS.415.3649V, 2019OJAp....2E...4C}, typically suppressing the matter power spectrum on scales of  $0.1\,h\,\mathrm{Mpc}^{-1} \lesssim k \lesssim 10\,h\,\mathrm{Mpc}^{-1}$ \citep{2023MNRAS.524.2539P, 2023MNRAS.526.4978S}. These baryonic feedback effects will introduce substantial systematic uncertainties into weak lensing measurement, and make a significant challenge for precise cosmological probes.

Since baryonic feedback is intrinsically linked to the spatial distribution of baryons, observational probes of the baryon distribution can provide valuable information on feedback processes. 
For example, the fast radio bursts (FRBs) is a promising probe for studying baryonic physics, which are extremely bright radio transients \citep{2007Sci...318..777L}. A key observable of FRBs is the dispersion measure (DM), which traces the integrated free electron density along the line of sight and therefore encodes both cosmological and baryonic information \citep{2014ApJ...780L..33M}. 

With the discovery of an increasing number of FRBs and the improvement in their precise localization, FRBs have become a more and more important probe in cosmology \citep[e.g.][]{2023RvMP...95c5005Z}. In particular, the DM-$z$ relation provides a powerful way to constrain cosmological parameters as well as the strength of baryonic feedback \citep{2020Natur.581..391M,2022MNRAS.516.4862J,2024arXiv240308611T,2024ApJ...973..151K,2025arXiv250717742R,2025ApJ...989...81S}.
Furthermore, FRBs can also serve as a tracer of the large--scale structure of the Universe and can be cross--correlated with other cosmological probes \citep{2017PhRvD..95h3012S,2020PhRvD.102b3528R,2021ApJ...922...42R,2022MNRAS.512.1730S,2025arXiv250919514L,2026ApJ...998..252C,2026arXiv260121336S}. 
Recently, several studies have reported detections of cross-correlation signals between FRBs and other tracers using existing FRB data \citep[see e.g.][]{2025arXiv251102155T,2025arXiv250608932W,2025ApJ...993L..27H,2026arXiv260422105S}. 

For next generation radio telescopes, such as the Square Kilometre Array (SKA; \citealp{2004NewAR..48..979C}) and the Deep Synoptic Array (DSA-2000; \citealp{2019BAAS...51g.255H}), the detection rate and localization precision of FRBs are expected to be improved substantially. As a result, the joint analysis of FRBs with other LSS tracers will become a powerful tool for studying the baryonic feedback effect and cosmological constraints \citep{2023arXiv230909766R,2026ApJ...998..109S,2026MNRAS.tmp..515W}.

In this work, we forecast the joint constraint power on baryonic feedback effect and cosmological parameters for the  CSST photometric survey together with future FRB measurements. CSST is a 2-meter space telescope designed for wide-field cosmological observations. CSST will conduct ten-year photometric imaging and slitless spectroscopic surveys covering approximately $17,500\,\mathrm{deg}^{2}$ with a field of view of $1.1\,\mathrm{deg}^{2}$, thereby enabling detailed mapping of the LSS of the Universe \citep{2026SCPMA..6939501C}. These observations will  significantly advance our understanding of the composition and evolution of the Universe \citep{2011SSPMA..41.1441Z,2019ApJ...883..203G,2022MNRAS.515.5743L,Zhan2021TheWM,2025SCPMA..6880402G,2025arXiv251122575H,2026ApJ...998..320X,2026ApJ..1000..143S,2026ApJ...997..357Y}.

For future FRB observations, we consider SKA1-Mid and DSA-2000 as representative next-generation radio facilities. Specifically, the SKA1-Mid provides full azimuthal coverage from $-270^\circ$ to $+270^\circ$ and operates over elevation angles ranging from $+15^\circ$ to $+95^\circ$. In addition, covering a frequency range of 350 MHz to 15.4 GHz, SKA1-Mid is expected to probe an area of approximately $30000\,\mathrm{deg}^{2}$ for FRB detection \citep{ska_lev0}. Similarly, DSA-2000 operates over a frequency range of 0.7--2.0 GHz and is designed to observe the entire sky above a declination of $-30^\circ$, thereby achieving an FRB survey footprint of about $30000\,\mathrm{deg}^{2}$ \citep{2019BAAS...51g.255H}.

The paper is structured as follows: in Section~\ref{sec:model}, we introduce the theoretical framework, including the modeling of power spectra; in Section~\ref{sec:mock data}, we show the methodology for generating mock data; in Section~\ref{sec:result}, we present and discuss the constraint results. In Section~\ref{sec:summary}, we summarize our findings and conclusions.

\section{Theoretical model}\label{sec:model}
\subsection{FRB DM statistics}
As FRBs propagate through the Universe, they are dispersed by ionized gas, causing signals at different frequencies to arrive at different times. This relative time delay is inversely proportional to the square of the frequency, $\Delta t \propto \nu^{-2}$, where the proportionality factor is defined as the dispersion measure. The total observed $\rm DM_{obs}$ along a given line of sight $\hat{\mathbf{n}}$ at redshift $z$ can be decomposed into three components:
\begin{equation}
    \mathrm{DM}_{\mathrm{obs}}(\mathbf{\hat{n}}, z) = \mathrm{DM}_{\mathrm{MW}}(\mathbf{\hat{n}}) + \mathrm{DM}_{\mathrm{LSS}}(\mathbf{\hat{n}}, z) + \frac{\mathrm{DM}_{\mathrm{host}}}{1 + z},
\end{equation}
where $\mathrm{DM}_{\mathrm{MW}}$ represents the contribution from the Milky Way, including both the interstellar medium (ISM) and circumgalactic medium (CGM). The term $\mathrm{DM}_{\mathrm{host}}$ denotes the contribution from the host galaxy in its rest frame, where the factor $(1+z)^{-1}$ accounts for the cosmological redshift. $\mathrm{DM}_{\mathrm{LSS}}$ arises from the LSS of the Universe, and it carries essential information regarding the cosmic web and can be expressed as
\begin{equation}
    \mathrm{DM}_{\mathrm{LSS}}(\mathbf{\hat{n}}, z) = \int_{0}^{z} \frac{\bar{n}_{e}(z') [1 + \delta_{e}(\mathbf{\hat{n}}, z')] (1 + z')}{H(z')} \, \mathrm{d}z'.
\end{equation}
Here $\delta_e$ denotes the electron density contrast, and $\bar{n}_e$ represents the mean comoving electron number density, which is given by
\begin{equation}
    \bar{n}_e(z) = \frac{f_{\mathrm{d}}(z) \bar\rho_{\mathrm{b}}(z) \chi_e(z)}{m_{\mathrm{p}}},
\end{equation}
where $f_{\rm d}$ is the fraction of baryons in the diffuse ionized gas. In this work, we adopt $f_{\rm d} = 0.93$, following the values derived from the Deep Synoptic Array (DSA-110) observational data as reported by \cite{2025NatAs...9.1226C}. The ionization fraction of electrons $\chi_{e}$ is given by $\chi_{e} \simeq Y_{\rm H} + \frac{1}{2} Y_{\rm He} \simeq 0.875$. Here, we assume that all electrons are fully ionized, which is a reasonable approximation given that the redshift distribution of FRBs is predominantly within $z < 3$. The $m_{\rm p}$ denotes the proton mass, and $\bar{\rho}_{\rm b} = \rho_{\rm crit}\Omega_{\rm b}(1+z)^3$ is the baryon mean density, where $\rho_{\rm crit} = 3H_0^2/(8\pi G)$ is the present-day critical density.
Assuming a spatially flat $\Lambda$CDM cosmology, $\rm{DM}_{\rm{LSS}}$ can be estimated by \citep{2023arXiv230909766R}
\begin{equation}
\begin{split}
\mathrm{DM}_{\mathrm{LSS}}(\mathbf{\hat{n}}, z) &= \int_{0}^{z} \frac{3c \chi_e \Omega_{\mathrm{b}} H_0}{8\pi G m_{\mathrm{p}}} \frac{f_{\mathrm{d}}(1+z') [1 + \delta_e(\mathbf{\hat{n}}, z')]}{\sqrt{\Omega_{\mathrm{m}}(1+z')^3 + \Omega_{\Lambda}}} \, \mathrm{d}z' \\
&= \langle {\rm DM}_{\rm LSS}(z) \rangle
 + \mathcal{D}(\hat{\mathbf{n}}, z).
\label{eq:DM_LSS}
\end{split}
\end{equation}
Here we decompose $\mathrm{DM}_{\mathrm{{LSS}}}(\hat{\mathbf{n}}, z)$ into a background contribution $\langle {\rm DM}_{\rm LSS}(z) \rangle$ and a perturbation term $\mathcal{D}(\hat{\mathbf{n}}, z)$. Considering the redshift distribution of FRBs, $n_{\rm f}(z)$, the average DM perturbation along a given line of sight $\hat{\mathbf{n}}$ can be written as
\begin{equation}
\mathcal{D}(\hat{\mathbf{n}}) = \int_{0}^{z_{\rm H}} \mathrm{d}z \, n_{\rm f}(z) \mathcal{D}(\hat{\mathbf{n}}, z),\label{eq:DM_perturbation}
\end{equation}
where $z_{\rm H} = 4$ denotes the maximum redshift of the FRB sample. By combining Eq.~\eqref{eq:DM_LSS} and Eq.~\eqref{eq:DM_perturbation}, the integrated perturbation $\mathcal{D}(\hat{\mathbf{n}})$ can be derived as
\begin{equation}
\mathcal{D}(\hat{\mathbf{n}}) = \int_0^{z_{\rm H}} \mathrm{d}z \, W_{\rm D}(z) \delta_e(\hat{\mathbf{n}}, z),
\end{equation}
where the radial kernel for the FRB DM field, $W_{\rm D}(z)$, is given by \citep{2026ApJ...998..109S}
\begin{equation}
W_{\rm D}(z) = \frac{\mathrm{d}\langle {\rm DM}_{\rm LSS}(z) \rangle
}{\mathrm{d}z} \int_z^{z_{\rm H}} \mathrm{d}z' n_{\rm f}(z').
\end{equation}

Since the physical origin and progenitor channels of FRBs remain uncertain, their intrinsic redshift distribution cannot yet be uniquely determined. We therefore adopt a physically motivated model in which the FRB source distribution follows the cosmic star formation rate (SFR) convolved with a log-normal delay-time distribution \citep{2008ApJ...683L...5Y,2015MNRAS.448.3026W}, following \cite{2021MNRAS.501..157Z} and \cite{2024ApJ...975..184W}. This prescription provides a phenomenological description consistent with several proposed stellar-origin FRB channels, including magnetars formed through binary white dwarf mergers, accretion-induced collapse of white dwarfs  and neutron star mergers \citep{2017ApJ...841...14M, 
2019ApJ...886..110M, 2021ApJ...907L..31B}. Moreover, recent CHIME/FRB population studies suggest that the cosmic evolution of the FRB population is delayed relative to the cosmic star formation history and can be well described by a log-normal delay-time model \citep{2024ApJ...969..123L,2026arXiv260704792Z}.

The intrinsic source redshift distribution of FRBs can be written as
\begin{equation}
n_{\text{s}}(z) \propto \frac{4\pi\chi^2(z) R(z)}{H(z)(1+z)},
\end{equation}
where $\chi(z)$ denotes the comoving radial distance, and $R(z)$ represents the event-rate evolution. {In this work, the event-rate evolution $R(z)$ is modeled as

\begin{equation}
\begin{aligned}
R(z) = & \left[ (1 + z)^{5.7\tau} + \left( \frac{1 + z}{0.36} \right)^{1.3\tau} + \left( \frac{1 + z}{3.3} \right)^{-9.5\tau} \right. \\ 
& \left. + \left( \frac{1 + z}{3.3} \right)^{-24.5\tau} \right]^{1/\tau},
\label{eq:FRB_redshift}
\end{aligned}
\end{equation}
where the parameter $\tau = -2$. This empirical form was adopted from \citet{2015ApJ...812...33S}, who used the constraints from \citet{2015MNRAS.448.3026W}, corresponding to a characteristic delay time of 2.9 Gyr and a logarithmic width of 0.2, to generate the redshift distribution as a log-normal delay-time model. They subsequently fitted the resulting distribution, yielding the form given in Eq.~(\ref{eq:FRB_redshift}).}

Due to the limited sensitivity of radio telescopes, intrinsically faint FRBs at high redshifts are difficult to detect. Therefore, the observed redshift distribution of FRBs must incorporate observational selection effects. The energy distribution of FRBs is commonly assumed to follow a functional form parameterized as

\begin{equation}
\frac{dN}{dE} \propto \left( \frac{E}{E_{\rm c}} \right)^{-\alpha} \exp\left( -\frac{E}{E_{\rm c}} \right), \label{eq:FRB_energy}
\end{equation}
where $E_{\rm c}$ is the exponential cut-off energy and $\alpha \simeq 1.8$ \citep{2020MNRAS.494..665L, 2020MNRAS.498.1973L}. Specifically, \citet{2020MNRAS.498.1973L} suggested a luminosity cutoff at $L_{\rm c} \sim 3 \times 10^{44} \text{ erg s}^{-1}$, which corresponds to a cut-off energy of $E_{\rm c} \sim 3 \times 10^{41} \text{ erg}$.

\begin{figure}[t!]
    \centering
	\includegraphics
    [width=\linewidth]{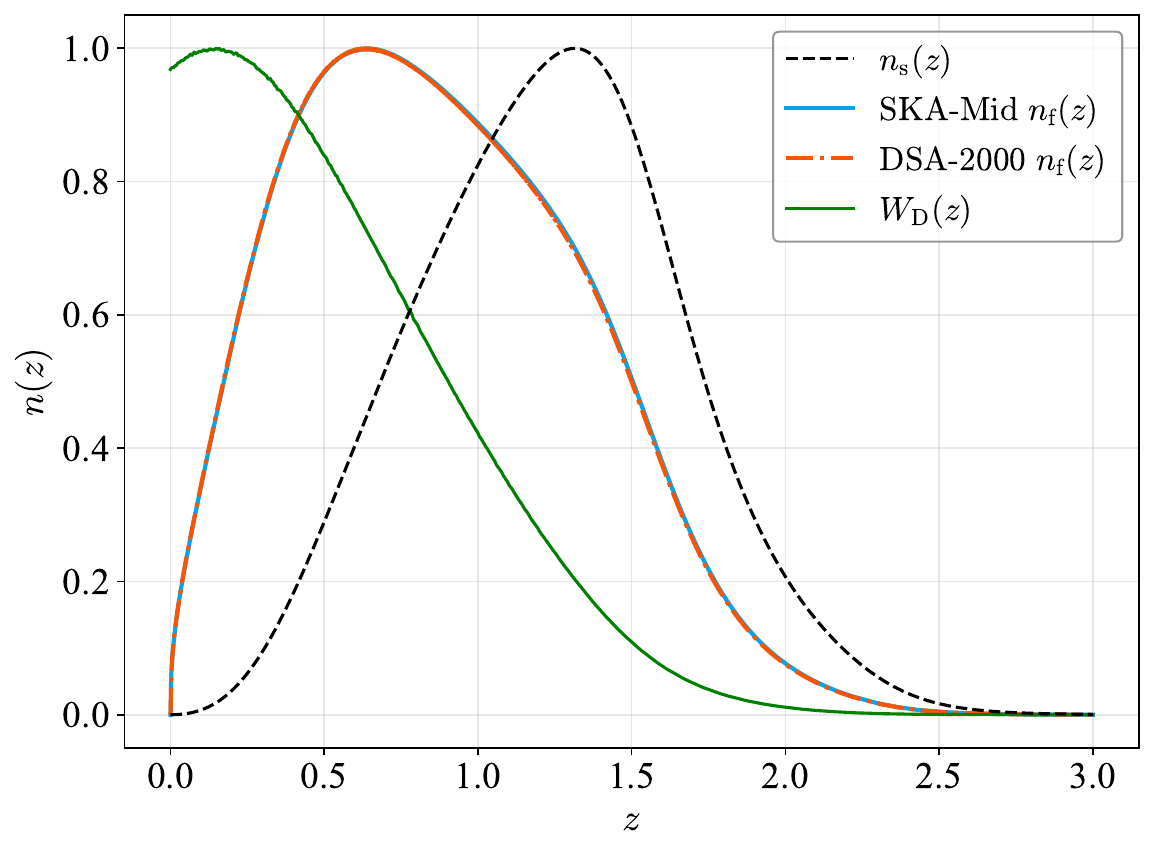}
    \caption{The rescaled intrinsic FRB source redshift distribution, $n_{\rm s}(z)$, observed redshift distribution after accounting for selection effects for SKA1-Mid and DSA-2000, $n_{\rm f}(z)$, and corresponding radial kernel of the FRB DM field, $W_{\rm D}(z)$.
}
    \label{fig:nz}
\end{figure}

Assuming a flat radio spectrum, the specific fluence $F_{\nu}$ detected from FRBs with isotropic energy $E$ at redshift $z$ can be written as \citep{2023RvMP...95c5005Z}
\begin{equation}
   F_{\nu} = \frac{(1 + z)E}{4\pi D_{\rm L}^{2}\Delta{\nu}},
\end{equation}
where $\Delta{\nu}$ is the bandwidth and $D_{\rm L}$ is the luminosity distance. 
For the FRB surveys with  SKA1-Mid and DSA-2000, 
we adopt the bandwidth of $\Delta{\nu} = 800\,{\rm MHz}$ 
and $1.3\,{\rm GHz}$, with $10\sigma$ fluence thresholds of 
$F_{\nu} = 14\,{\rm mJy\,ms}$ 
and $15.5\,{\rm mJy\,ms}$, respectively \citep{ska_lev0,2015aska.confE..51F,2019BAAS...51g.255H}.
We simulate $10^{8}$ FRBs whose redshift and energy distributions follow Eq.~\eqref{eq:FRB_redshift} and Eq.~\eqref{eq:FRB_energy}, respectively. This large sample ensures a stable reconstruction of the observable redshift distribution after applying the fluence selection. The fluence of each FRB is then computed, and only those with fluence exceeding the telescope sensitivity threshold are retained. Finally, we derive the expected redshift distribution of the observable FRBs, $n_{\rm f}(z)$, along with the corresponding radial kernel for the FRB DM field, $W_{\rm D}(z)$, as shown in Figure~\ref{fig:nz}.
As can be seen, since the detection capabilities of DSA-2000 and SKA1-Mid are similar, we adopt the instrumental specifications of SKA1-Mid in our calculations as representative of next-generation FRB surveys.

As for the expected number of detected FRBs, $N_{\rm FRB}$, 
we adopt a power-law model as a reference estimate. 
The all-sky FRB event rate $N_{\rm sky}$ above a fluence threshold $F_{\nu}$ 
is assumed as \citep{2018MNRAS.480.4211M}
\begin{equation}
N_{\rm sky}(>F_{\nu}) = N'_{\rm sky}
\left(\frac{F_{\nu}}{F'_{\nu}}\right)^{\lambda},
\end{equation}
where the power-law index $\lambda = -1.5$, as expected for a non-evolving population in Euclidean space.  This relation follows from the inverse-square scaling of fluence, $F_{\nu}\propto r^{-2}$, and the volume scaling of detectable sources, $N_{\rm sky}\propto r_{\rm max}^{3}$, which together lead to $N_{\rm sky}(>F_{\nu})\propto F_{\nu}^{-3/2}$. Based on the results from the Parkes and ASKAP FRB surveys, the all-sky rate for SKA1-Mid is estimated to be
$
N_{\rm sky} \simeq 3 \times 10^{6} \ {\rm sky^{-1}\,day^{-1}} 
$
\citep{2023SCPMA..6620412Z}.
The FRB detection rate can then be estimated as
$
N_{\rm FRB} =
N_{\rm sky} \,
T \,
\Delta \Omega/(4\pi),
$
where $T$ is the total observation time and $\Delta \Omega$ is the telescope 
field of view.
For $\rm SKA1\mbox{-}Mid$, which has an effective instantaneous wide field of 
view of $\sim 20\,{\rm deg^{2}} $ \citep{2010PASA...27..272M}, and assuming that $20\%$ of the observing 
time in a year is used for FRB searches, the expected detection rate is 
about $10^{5}\,{\rm yr^{-1}}$ \citep{2023SCPMA..6620412Z}. Therefore, we set $N_{\rm FRB} = 10^{5}$ as the fiducial number of FRBs used in the analysis.

Under the Limber and flat-sky approximation, the angular power spectrum of the DM perturbation field, $C_{\rm DD}$, can be expressed as
\begin{equation}
C_{\rm DD} = \int_0^{\chi_{\rm H}} \mathrm{d}\chi \frac{W_{\rm D}^2(\chi)}{r^2(\chi)} P_{ee} \left( k = \frac{\ell + 1/2}{r(\chi)}, z \right),
\end{equation}
where $r(\chi)$ is the comoving angular diameter distance and $P_{ee}$ is the electron power spectrum. 

To account for the impact of baryonic physics on angular power spectra, we employ the halo model framework to construct the electron power spectrum $P_{ee}$, the matter power spectrum $P_{\rm mm}$, and matter-electron cross-power spectrum $P_{{\rm m}e}$ . In this approach, the statistical distribution of matter is determined by the properties of individual halos and their spatial correlations. We adopt the baryonic feedback model proposed by \citet{2020A&A...641A.130M} in this work. Within this framework, the total matter content of halos is decomposed into three components: cold dark matter, gas, and stars. The model provides a parameterized description of both the mass fractions and the density profiles of these components. The corresponding mass fractions depend on the total halo mass $M$ and are denoted as $f_{\rm c}(M)$, $f_{\rm g}(M)$, and $f_{*}(M)$, respectively, satisfying
$
f_{\rm c}(M) + f_{\rm g}(M) + f_{*}(M) = 1.
$

\begin{figure*}[t!]
    \centering
    \includegraphics[width=1\textwidth]{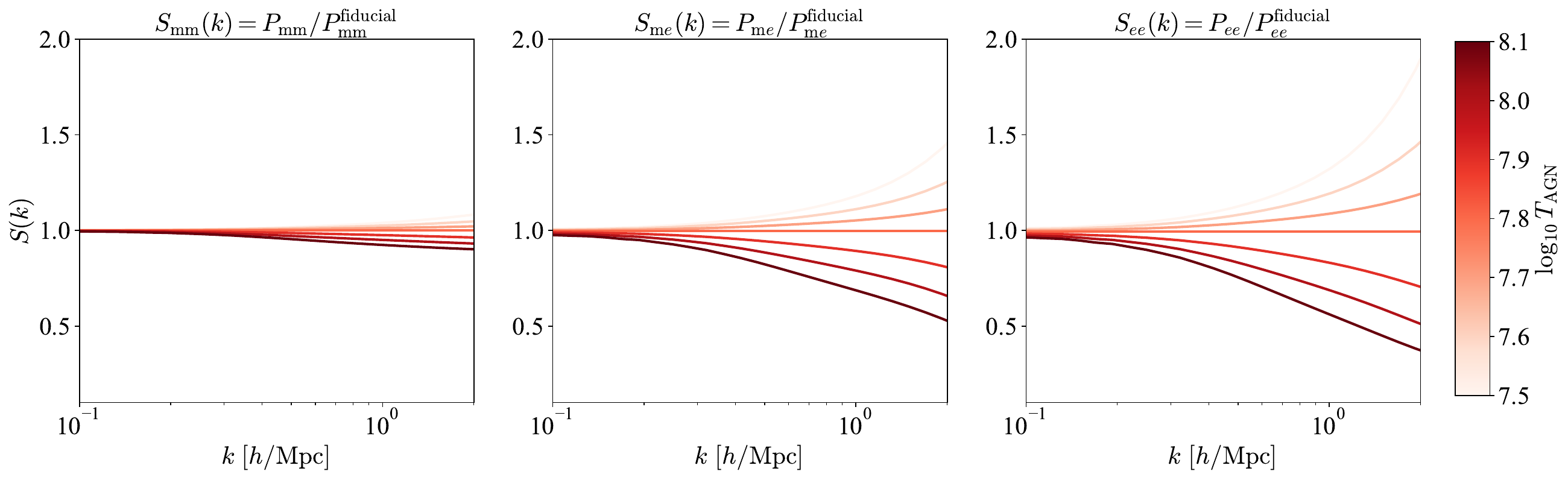}
    \caption{Impact of baryonic feedback parameterized by $T_{\rm AGN}$ on $P_{\rm mm}$, $P_{{\rm m}e}$, and $P_{ee}$ at redshift $z = 0$. Each power spectrum computed with a different $T_{\rm AGN}$ is divided by the fiducial spectrum with $\log_{10} T_{\rm AGN} = 7.8$ to illustrate the sensitivity to baryonic feedback strength.
}\label{fig:sk}
\end{figure*}

The fraction of cold dark matter follows the cosmic value and can be written as
$
f_{\rm c}(M) = \Omega_{\rm c}/\Omega_{\rm m},
$
where $\Omega_{\rm c}$ and $\Omega_{\rm m}$ denote the cosmic density parameters of cold dark matter and total matter, respectively. 
Hydrodynamical simulations suggest that baryonic feedback can modify the inner structure of dark matter halos. Nevertheless, the cold dark matter density profile can still be well approximated by the Navarro-Frenk-White (NFW) form \citep{1997ApJ...490..493N,  2020A&A...641A.130M}. To account for baryonic effects, the concentration parameter $c(M)$ is modified as
\begin{equation}
c(M) \rightarrow
c(M)\left[
1 + \epsilon_1
+ (\epsilon_2 - \epsilon_1)
\frac{f_{\mathrm{bond}}(M)}{\Omega_{\mathrm{b}}/\Omega_{\mathrm{m}}}
\right],
\end{equation}
where $\epsilon_1$ and $\epsilon_2$ are the parameters characterizing the impact of baryonic feedback on halo concentrations.

For the gas component, the total gas fraction is further divided into bound gas and ejected gas. The bound gas refers to the gas confined within the virial radius of the halo due to gravitational attraction. Its fraction, $f_{\rm bond}(M)$, follows the functional form proposed by \citet{2015JCAP...12..049S}:
\begin{equation}
    f_{\rm bond}(M) =
    \frac{\Omega_{\rm b}}{\Omega_{\rm m}}
    \frac{(M/M_{0})^{\beta}}
    {1+(M/M_{0})^{\beta}},
\end{equation}
where $\beta$ controls the mass dependence of the transition in the bound gas fraction and $M_{0}$ represents the characteristic halo mass at which half of the baryons remain bound within the halo.
The density profile of the bound gas follows the functional form proposed by \citet{2013MNRAS.432.1947M}:
\begin{equation}
    \rho_{\mathrm{bond}}(M,r)
    \propto
    \left[
    \frac{\ln(1+r/r_{\mathrm{s}})}
    {r/r_{\mathrm{s}}}
    \right]^{1/(\Gamma-1)},
\end{equation}
where $\Gamma$ denotes the polytropic index of the gas profile. 

The fraction of ejected gas is defined as
\begin{equation}
    f_{\rm ejc}(M)
    =
    \frac{\Omega_{\rm b}}{\Omega_{\rm m}}
    -
    f_{\rm bond}(M)
    -
    f_{*}(M).
\end{equation}
The spatial distribution of the ejected gas is modeled following \citet{2014JCAP...04..028F} and \citet{2020MNRAS.492.2285D}. 
The profile of the ejected gas is treated as a delta function in the two-halo term, while its contribution to the one-halo term is neglected. This approximation assumes that the ejected gas contributes only to large-scale clustering and is ignored on small scales \citep{2020A&A...641A.130M}. 

The stellar fraction, $f_{\ast}(M)$, takes the form given by \cite{2014JCAP...04..028F}:
\begin{equation}
    f_{\ast}(M) = A_{\ast} \exp \left[ -\frac{\log_{10}^{2}(M/M_{\ast})}{2\sigma_{\ast}^{2}} \right],
\end{equation}
where the stellar fraction reaches its maximum value $A_{\ast}$ at a halo mass of $M_{\ast}$, and $\sigma_{\ast}$ characterizes the width of this distribution, representing the range of halo masses over which star formation is efficient. 
The stellar fraction is further decomposed into two components: the contribution from the central galaxy and that from satellite galaxies. For halos with $M < M_{*}$, we have
\begin{equation}
f_{\rm cen}(M) = f_{*}(M), \quad f_{\rm sat}(M) = 0,
\end{equation}
assuming that all stellar mass resides in the central galaxy. 
For more massive halos with $M > M_{*}$, we have
\begin{equation}
f_{\rm cen}(M) = f_{*}(M)\left(\frac{M}{M_{*}}\right)^{\eta},
\end{equation}
\begin{equation}
f_{\rm sat}(M) = f_{*}(M)\left[1 - \left(\frac{M}{M_{*}}\right)^{\eta}\right],
\end{equation}
where $\eta$ controls the slope of the dependence of the central galaxy mass on the stellar mass. As for their profiles, the central galaxy is modeled as a delta function, while the satellite galaxies are assumed to follow the NFW profile \citep{2020A&A...641A.130M}.

The baryonic feedback model includes the parameters $\left\{\epsilon_1, \epsilon_2, \beta, M_{0}, \Gamma, A_{\ast}, M_{\ast}, \sigma_{\ast}, \eta \right\}$. Among them, $\epsilon_2$, $\beta$, and $\sigma_{\ast}$ are fixed to $0$, $0.6$, and $1.2$, respectively, while the remaining six parameters $\left\{{\epsilon_1, M_{0}, \Gamma, A_{\ast}, M_{\ast}, \eta}\right\}$  are treated as free parameters \citep{2020A&A...641A.130M}. The model is calibrated against the AGN BAHAMAS hydrodynamical simulations \citep{2017MNRAS.465.2936M} spanning a range of AGN heating temperatures, $T_{\rm AGN}$. By establishing a mapping between the model parameters and the corresponding heating temperatures, the overall strength of baryonic feedback can be effectively characterized by the single parameter $T_{\rm AGN}$.

We further assume that the gas is fully ionized such that the electron distribution traces the gas distribution. Under this assumption, the electron power spectrum follows that of the gas component, which is a valid approximation \citep{2025arXiv250919514L}.
Within this framework, we compute the electron power spectrum $P_{ee}$, the matter power spectrum $P_{\rm mm}$,  and their cross-power spectrum $P_{{\rm m}e}$ for different baryonic feedback strengths using the \texttt{pyhmcode} \footnote{\url{https://github.com/tilmantroester/pyhmcode?tab=readme-ov-file}} \citep{2020A&A...641A.130M,2022A&A...660A..27T} and \texttt{pyccl} \footnote{\url{https://github.com/LSSTDESC/CCL?tab=readme-ov-file}} \citep{2019ApJS..242....2C} packages.

In Figure~\ref{fig:sk}, we show the ratios of the power spectra at $z=0$ computed for different $T_{\rm AGN}$ values relative to the fiducial model with $\log_{10} T_{\rm AGN}=7.8$. With increasing baryonic feedback strength, the power spectra are suppressed mainly on small scales. This is caused by the redistribution of baryonic components, leading to a reduction of the one-halo term. In contrast, the two-halo term, which dominates large-scale clustering, remains relatively insensitive to the variation of baryonic feedback.
We can also find that both the electron power spectrum and the matter-electron cross-power spectrum exhibit significantly higher sensitivity to baryonic feedback effect than the matter power spectrum. 
Consequently, the FRB DM field can serve as a powerful probe of baryonic feedback processes. Since baryonic feedback introduces significant systematics in weak lensing measurements at small scales, combining FRB DM observations with weak lensing provides an effective approach for constraining both baryonic physics and cosmological parameters.

Additionally, since the contributions from $\mathrm{DM}_{\rm MW}$ and $\mathrm{DM}_{\rm host}$ must be removed in order to infer $\mathrm{DM}_{\rm LSS}$, we need to account for the impact of uncertainties associated with these components on our results.
For $\mathrm{DM}_{\rm MW}$, acting as a foreground, it has been accurately modeled \citep{2026arXiv260211838O,2017ApJ...835...29Y,2002astro.ph..7156C}.
In particular, this foreground contribution exhibits relatively large uncertainties at low Galactic latitudes, whereas in high Galactic latitude regions the associated uncertainties are sufficiently small to be effectively removed. Since our analysis only considers the overlapping sky region between the weak lensing survey and FRB observations, which lies predominantly at relatively high Galactic latitudes, we neglect this term in this study.
In contrast, the contribution from the host galaxy is more difficult to model precisely. 
{The host-galaxy DM contribution can in principle exhibit spatial correlations, with the amplitude of the correlated component estimated to be below the 10\% level relative to the $\rm DM_{LSS}$ signal \citep{2017PhRvD..95h3012S,2026MNRAS.tmp..515W}. Here, we model the residual host-galaxy contribution as an effective stochastic scatter \citep{2023arXiv230909766R, 2026ApJ...998..109S}, which serves as a simplified  approximation in the analysis.}
To account for the intrinsic scatter of the host galaxy contribution, we adopt a characteristic uncertainty of $\sigma_{\rm host,0} = 90 \, \rm pc \, cm^{-3}$, which represents a reasonable assumption consistent with current observational estimates \citep{2026ApJ...998..109S,2026MNRAS.tmp..515W}. This uncertainty introduces an additional noise contribution to the FRB DM field. 

\begin{figure}[t!]
	\centering
	\includegraphics[width=\linewidth]{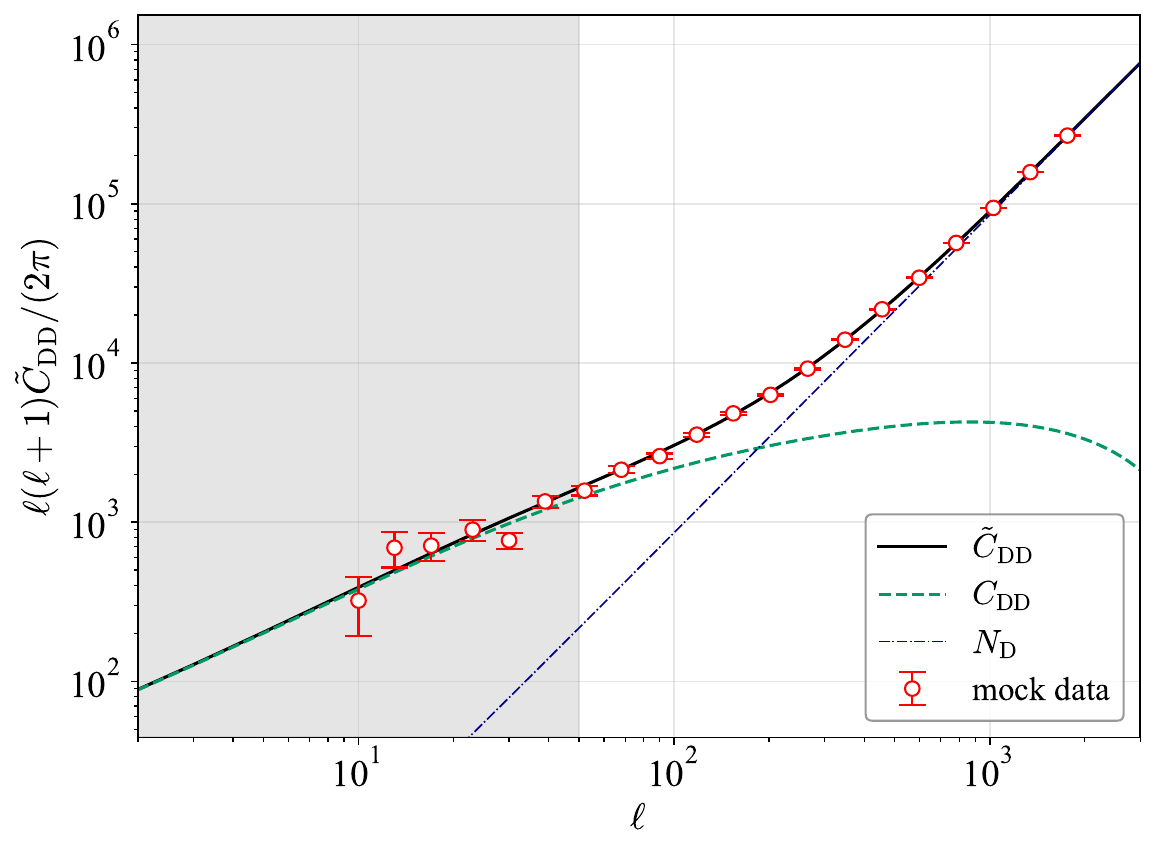}
    \caption{The theoretical predictions and mock data for the angular power spectrum of the FRB DM field. 
    The black solid curve shows the fiducial theoretical model, $\tilde{C}_{\rm DD}=C_{\rm DD}+N_{\rm D}$, while the green dashed and blue dash-dotted curves represent the signal term, $C_{\rm DD}$, and the noise term, $N_{\rm D}$, respectively.
    The red data points denote the mock data used in this work. The data points in the shaded region at $\ell<50$ are excluded from the analysis, where the flat-sky and Limber approximations are not available.} 
	\label{fig:data_ff}
\end{figure}

\begin{figure*}[t!]
    \centering
    \includegraphics
    [width=\linewidth]{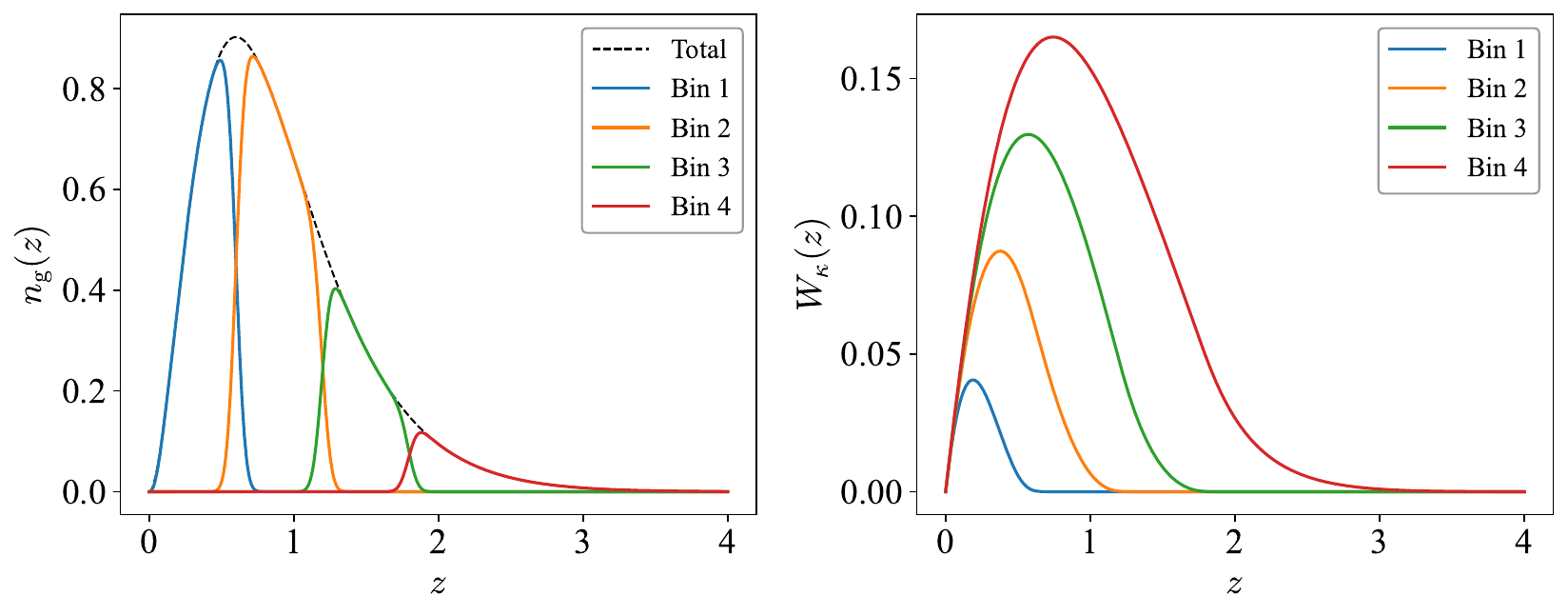}
    \caption{The CSST redshift distributions of galaxies (left panel), $n_{\rm g}^{i}(z)$, and the weak lensing kernel (right panel), $W_{\kappa}^{i}(z)$, for the $i$-th tomographic bin.
}
    \label{fig:nz_galaxy}
\end{figure*}

Consequently, the total noise term of the DM field, $N_{\rm D}$, can be expressed as
\begin{equation}
 N_{\rm D} = \frac{\sigma_{\rm D}^2 + \sigma_{\rm host}^2}{\bar{n}_{\rm FRB}},
\end{equation}
where $\bar{n}_{\rm FRB}$ represents the average surface number density of FRBs.  Here,
\begin{equation}
\sigma^2_{\rm host} = \int dz\, n_{\rm f}(z)\, \frac{\sigma^2_{\rm host,0}}{(1+z)^2},
\end{equation}
and 
\begin{equation}
\sigma_{\rm D}^2 = \sum_{\ell} \frac{2\ell + 1}{4\pi}\, C_{\rm DD}(\ell),
\end{equation}
which describes the variance contributed by the stochastic DM field fluctuations arising from different lines of sight \citep{2026MNRAS.tmp..515W}. Consequently, the observed angular power spectrum is given by
\begin{equation}
\tilde{C}_{\rm DD} = C_{\rm DD} + N_{\rm D}.
\end{equation}
When generating the mock data, the noise term $N_{\rm D}$ is estimated using the fiducial parameter values and survey specifications. To account for uncertainties in the noise modeling, $N_{\rm D}$ is treated as a free nuisance parameter and marginalized over in the fitting process.
The resulting signal, noise, and total DM angular power spectra are shown in Figure~\ref{fig:data_ff}.

\begin{figure*}[t!]
    \centering
    \includegraphics[width=1\textwidth]{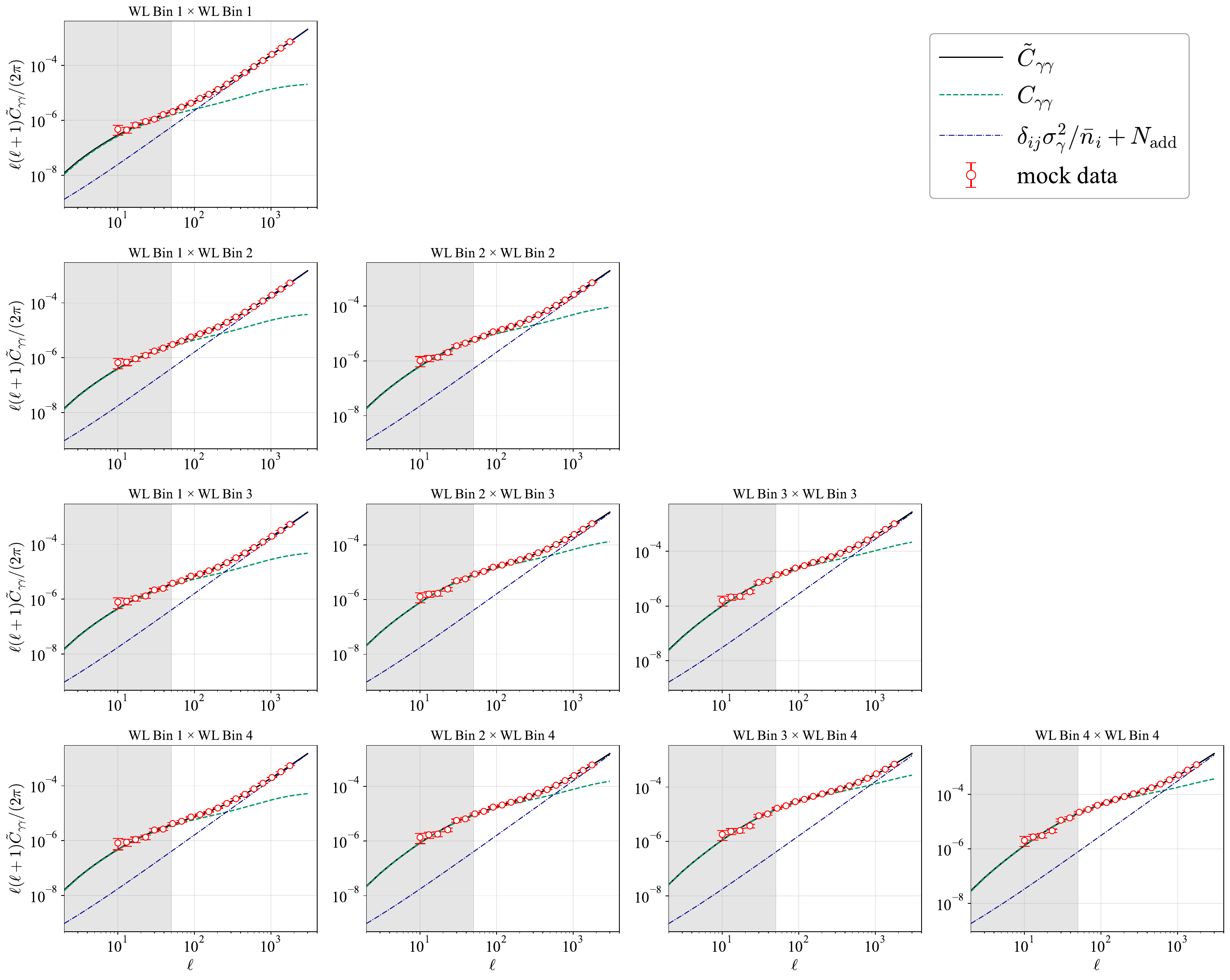}
    \caption{The theoretical predictions and mock data for the angular power spectra of weak lensing between the $i$-th and $j$-th tomographic bins, $\tilde{C}_{\gamma\gamma}^{ij}$. The black solid curve represents the fiducial theoretical model, while the green dashed curve shows the signal component, $C^{ij}_{\gamma\gamma}$, and the blue dash-dotted curve denotes the noise term, $\delta_{ij}\sigma_{\gamma}^2/\bar{n}_i + N_{\rm add}$. The red data points correspond to the mock data used in this work. The data points in the shaded region at $\ell<50$ are excluded from the analysis, where the flat-sky and Limber approximations are not available.}
    \label{fig:data_ww}
\end{figure*}

\subsection{Weak lensing}
The observed shear power spectrum $\tilde{C}_{\gamma \gamma}^{ij}(\ell)$ for the $i$-th and $j$-th tomographic bins at a given multipole $\ell$ can be written as \citep{2006MNRAS.366..101H,2008MNRAS.391..228A}
\begin{equation}
\tilde{C}_{\gamma \gamma}^{ij}(\ell) = (1 + m_i)(1 + m_j)C_{\gamma \gamma}^{ij}(\ell) + \delta_{ij}\frac{\sigma_{\gamma}^2}{\bar{n}_i} + N_{\rm add},
\end{equation}
where the multiplicative bias parameters $m_i$ and $m_j$ are introduced for each tomographic bin to account for systematic uncertainties in cosmic shear estimation. The statistical uncertainty is characterized by the shape noise term, $\delta_{ij}\sigma_{\gamma}^2/\bar{n}_i$, where $\sigma_{\gamma}=0.2$ denotes the root-mean-square intrinsic ellipticity per component and $\bar{n}_i$ is the mean surface number density of galaxies in the $i$-th tomographic bin. In addition, the term $N_{\rm add}$ accounts for residual additive shear biases arising from instrumental effects or imperfect point spread function modeling.
The theoretical cosmic shear power spectrum, $C_{\gamma \gamma}^{ij}(\ell)$, consists of contributions from gravitational lensing and intrinsic alignments (IA), and can be expressed by
\begin{equation}
C_{\gamma \gamma}^{ij}(\ell) = P_{\kappa}^{ij}(\ell) + C_{\rm II}^{ij}(\ell) + C_{\rm GI}^{ij}(\ell),
\end{equation}
where $P_{\kappa}^{ij}(\ell)$ represents the convergence power spectrum corresponding to the gravitational lensing signal. The remaining terms describe intrinsic alignment effects: $C_{\rm II}^{ij}(\ell)$ is the intrinsic-intrinsic power spectrum, which characterizes correlations between the intrinsic shapes of physically neighboring galaxies, while $C_{\rm GI}^{ij}(\ell)$ is the gravitational-intrinsic power spectrum, describing correlations between the intrinsic shape of a foreground galaxy and the lensing shear induced by a background galaxy \citep{2015SSRv..193....1J}.

Under the Limber and flat-sky approximation, the convergence power spectrum $P_{\kappa}^{ij}(\ell)$ can be derived as
\begin{equation}
P_{\kappa}^{ij}(\ell) = \int_0^{\chi_{\rm H}} \mathrm{d}\chi \, \frac{W_{\kappa}^i(\chi) W_{\kappa}^j(\chi)}{r^2(\chi)} \, P_{\rm mm}\left(k = \frac{\ell + 1/2}{r(\chi)}, z\right),
\end{equation}
where $W_{\kappa}^i(\chi)$ is the lensing kernel for the $i$-th tomographic bin, defined as
\begin{equation}
W_{\kappa}^i(\chi) = \frac{3\Omega_{\rm m} H_0^2}{2c^2} \frac{r(\chi)}{a(\chi)} \int_{\chi}^{\chi_{\rm H}} \mathrm{d}\chi' \, n^i_{\rm g}(\chi') \frac{r(\chi' - \chi)}{r(\chi')}.
\end{equation}
Here, $a(\chi)$ is the scale factor, and $n^i_{\rm g}(\chi)$ represents the normalized galaxy distribution of the $i$-th tomographic bin, satisfying $\int n^i_{\rm g}(\chi) \mathrm{d}\chi = 1$.

The overall redshift distribution of galaxies detected by the CSST photometric survey is given by \citep{2019ApJ...883..203G,2022MNRAS.515.5743L}
\begin{equation}
n_{\rm g}(z) \propto z^2 \exp(-z/z_*),
\end{equation}
where $z_* = z_{\rm peak}/2$, and $z_{\rm peak} = 0.6$ denotes the redshift at which the galaxy distribution reaches its maximum. The total galaxy sample is divided into four photometric tomographic bins. The mean galaxy surface number densities in these bins are $\bar{n}^{i}_{\rm g} = 7.9, 11.5, 4.6,$ and $3.7\ \mathrm{arcmin}^{-2}$, respectively, corresponding to a total number density of $\bar{n}_{\rm g} = 27.7\,\mathrm{arcmin}^{-2}$.

For the observed photometric redshift $z_{\rm p}$ of galaxies, we assume a Gaussian distribution
$
z_{\rm p} \sim \mathcal{N}(z_{\rm true},\, \sigma_z),
$
where $z_{\rm true}$ denotes the true redshift and $\sigma_z$ characterizes the photometric redshift uncertainty. To account for possible systematic uncertainties in the photometric redshift estimation, we introduce a shift parameter $\Delta z_i$ and a stretch parameter $\sigma_z^i$ for each tomographic bin. The redshift distribution in the $i$-th bin is  written as
\begin{equation}
n_{\rm g}^i(z) \rightarrow
\frac{1}{\sigma_z^i}
n_{\rm g}^i\!\left(
\frac{z-\bar{z}_i-\Delta z_i}{\sigma_z^i}
\right)
\end{equation}
where $\bar{z}_i$ is the mean redshift of the $i$-th tomographic bin.
Following \citet{2022MNRAS.515.5743L}, we adopt fiducial values of $\Delta z_i = 0$, $\sigma_z^i = 0.05$, multiplicative bias parameters $m_i = 0$, and additive bias $N_{\rm add} = 10^{-9}$, while treating them as free parameters in the fitting process. 
The corresponding galaxy redshift distributions $n_{\rm g}^{i}(z)$ and lensing kernels $W_{\kappa}^{i}(z)$ are shown in Figure~\ref{fig:nz_galaxy}.

\begin{figure*}[t!]
    \centering
    \includegraphics[width=1\textwidth]{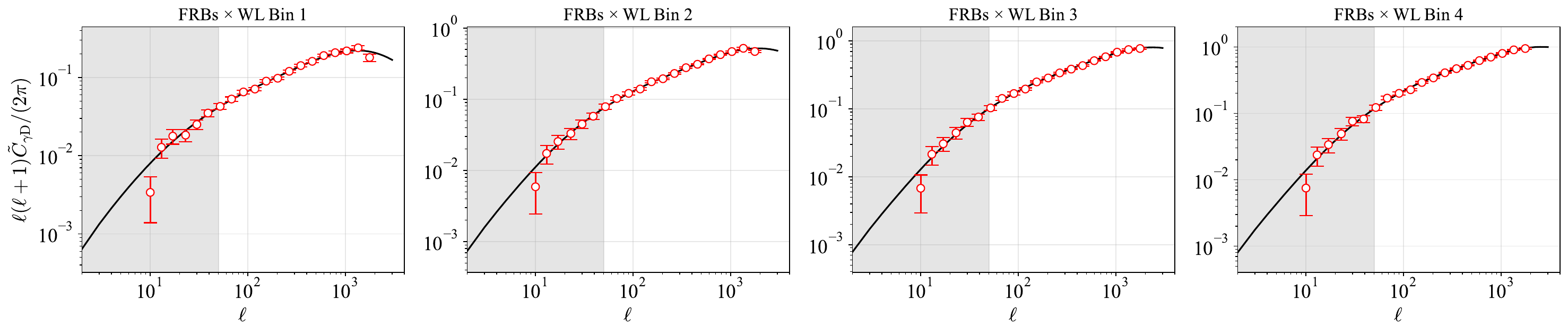}
    \caption{The theoretical predictions and mock data for the angular cross-power spectra between the FRB DM field and weak lensing in each tomographic bin, $\tilde{C}_{\gamma {\rm D}}^{i}$. The black solid curves represent the fiducial theoretical model, while the red data points denote the mock data used in this work. The data points in the shaded region at $\ell<50$ are excluded from the analysis, where the flat-sky and Limber approximations are not available.}
    \label{fig:data_fw}
\end{figure*}

Furthermore, the intrinsic-intrinsic and gravitational-intrinsic power spectra can be expressed as
\begin{equation}
C_{\rm II}^{ij}(\ell) = \int_{0}^{\chi_{\rm H}} \mathrm{d}\chi \, \frac{n_i(\chi)n_j(\chi)F_i(\chi)F_j(\chi)}{r^2(\chi)} P_{\rm mm}\left(\frac{\ell + 1/2}{r(\chi)}, z\right),
\end{equation}
and
\begin{equation}
\begin{split}
C_{\rm GI}^{ij}(\ell) &=  \int_0^{\chi_{\rm H}} \mathrm{d}\chi \frac{W_{\kappa}^i(\chi) n_j(\chi) F_j(\chi)}{r^2(\chi)} P_{\rm mm}\left(\frac{\ell + 1/2}{r(\chi)}, z\right) \\
& + \int_0^{\chi_{\rm H}} \mathrm{d}\chi \frac{W_{\kappa}^j(\chi) n_i(\chi) F_i(\chi)}{r^2(\chi)} P_{\rm mm}\left(\frac{\ell + 1/2}{r(\chi)}, z\right).
\end{split}
\end{equation}
The alignment factor $F(\chi)$ for the $i$-th tomographic bin is given by
\begin{equation}
F_i(\chi) = A_{\rm IA} C_1 \rho_{\rm crit} \frac{\Omega_{\rm m}}{D(\chi)} \left( \frac{1+z}{1+z_0} \right)^{\eta_{\rm IA}} \left( \frac{L_i}{L_0} \right)^{\beta_{\rm IA}},
\end{equation}
where $C_1 = 5 \times 10^{-14} \, h^{-2} M_{\odot}^{-1} \rm Mpc^3$ serves as the normalization constant, and the linear growth factor, $D(\chi)$, is normalized to unity at $z = 0$. The term $L_i$ represents the weighted average luminosity within the $i$-th tomographic bin, while $L_0$ is defined as the pivot luminosity. The pivot redshift $z_0$ denotes the reference redshift of the intrinsic alignment model, $A_{\rm IA}$ represents the alignment amplitude, $\eta_{\rm IA}$ is the power-law index describing the redshift evolution, and $\beta_{\rm IA}$ characterizes the luminosity dependence. 
Typically, $\beta_{\rm IA}$ is generally fixed to $0$ to neglect any potential luminosity-dependent scaling \citep{2017MNRAS.465.1454H,2018MNRAS.474.4894J}. 
In this work, we set $z_0=0.6$, treat $A_{\rm IA}$ and $\eta_{\rm IA}$ as free parameters with fiducial values of $1$ and $0$, respectively, and fix $\beta_{\rm IA}=0$ \citep{2019ApJ...883..203G}. The theoretical predictions for $\tilde{C}_{\gamma\gamma}^{ij}(\ell)$, together with the signal component $C_{\gamma\gamma}^{ij}(\ell)$ and noise term computed using the fiducial parameter values, are shown in Figure~\ref{fig:data_ww}.

\subsection{Cross-power spectrum}
The observed angular cross-power spectrum between the FRB DM and weak lensing for the $i$-th tomographic bin, $\tilde{C}_{\gamma \rm D}^{i}(\ell)$, can be derived as 
\begin{equation}
    \tilde{C}_{\gamma \rm D}^{i}(\ell) = C_{\kappa \rm D}^{i}(\ell) + C_{\rm I D}^{i}(\ell).
\end{equation}
Here, $C_{\kappa \rm D}^{i}(\ell)$ and $C_{\rm I D}^{i}(\ell)$ denote the cross-power spectra of the FRB DM with the convergence and the galaxy intrinsic alignment field, respectively. For simplicity, we set the noise term of the cross-power spectrum to be zero, since the noise of the two probes is essentially independent. Under the Limber and flat-sky approximation, they can be expressed as \citep{2025JCAP...12..035B}
\begin{equation}
C_{\rm \kappa D}^{i}(\ell) = \int_{0}^{\chi_{\rm H}} \mathrm{d}\chi \, \frac{W_{\kappa}^{i}(\chi)W_{\rm D}(\chi)}{r^2(\chi)} P_{{\rm m}e}\left(\frac{\ell + 1/2}{r(\chi)}, z\right)
\end{equation}
and
\begin{equation}
C_{\rm ID}^{i}(\ell) = \int_{0}^{\chi_{\rm H}} \mathrm{d}\chi \, \frac{n_i(\chi)F_i(\chi)W_{\rm D}(\chi)}{r^2(\chi)} P_{{\rm m}e}\left(\frac{\ell + 1/2}{r(\chi)}, z\right).
\end{equation}
The corresponding theoretical predictions for $\tilde{C}_{\rm \gamma D}^{i}(\ell)$ are shown in Figure~\ref{fig:data_fw}.

\section{Mock Data}\label{sec:mock data}
We adopt the Gaussian covariance matrix to describe the correlations and uncertainties among 
$\tilde{C}_{\rm DD}$, $\tilde{C}_{\gamma\gamma}$, and $\tilde{C}_{\gamma \rm D}$:
\begin{equation}
\begin{aligned}
&\mathrm{Cov}_{\rm ABEF}^{ijmn}\!\left[\tilde{C}^{ij}_{\rm AB}(\ell), \tilde{C}^{mn}_{\rm EF}(\ell')\right] \\
&= \frac{\delta_{\ell\ell'}}{(2\ell + 1)\, f_{\rm sky}\,\Delta\ell}
\left[
\tilde{C}^{im}_{\rm AE}(\ell)\tilde{C}^{jn}_{\rm BF}(\ell')
+
\tilde{C}^{in}_{\rm AF}(\ell)\tilde{C}^{jm}_{\rm BE}(\ell')
\right],
\end{aligned}
\end{equation}
where $\rm A,B,E,F \in \{\gamma, D\}$ denote the observed fields, and 
$i,j,m,n$ label different tomographic bins.
The parameter $f_{\rm sky}$ represents the observed sky fraction. 
In this work, we adopt $f_{\rm sky}=0.3$, corresponding to the overlapping survey areas 
between CSST and SKA1-Mid, as well as between CSST and DSA-2000 \citep{ska_lev0, 2019BAAS...51g.255H,2026SCPMA..6939501C}.

\begin{figure}[t!]
	\centering
	\includegraphics[width=\linewidth]{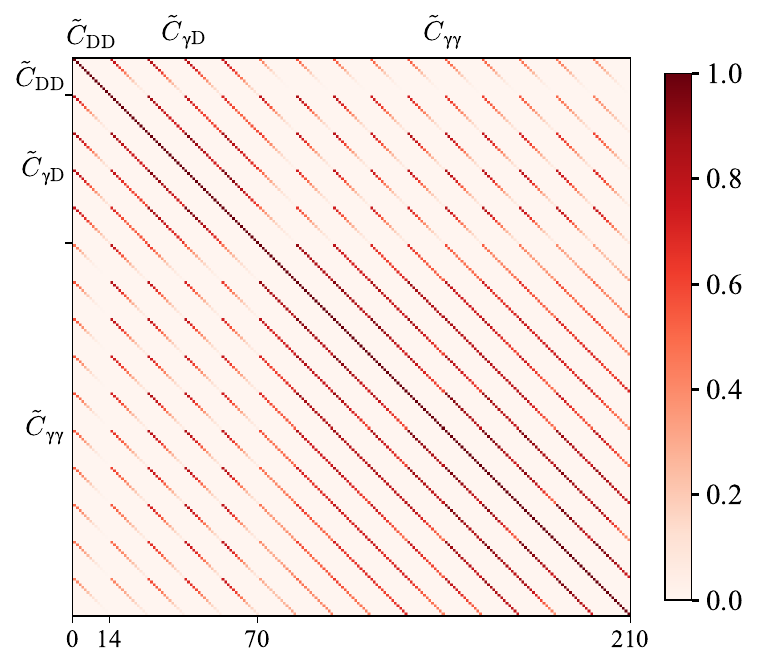}

    \caption{The covariance matrix for the FRB DM auto-power spectrum $\tilde{C}_{\rm DD}$, the weak lensing auto-power spectra $\tilde{C}_{\gamma\gamma}$, and their cross-power spectra $\tilde{C}_{\gamma {\rm D}}$, where the correlation coefficients are defined as $\tilde{C}_{ij}/\sqrt{\tilde{C}_{ii}\tilde{C}_{jj}}$, with $i$ and $j$ denoting the indices of the data points.}

	\label{fig:cov}
\end{figure}

To generate mock data, we first perform a Cholesky decomposition of the covariance matrix. 
The resulting lower-triangular matrix is then multiplied by a standard normal random matrix to produce 
random realizations. 
These random fluctuations are subsequently added to the theoretical predictions to construct the mock data.
We take $\ell_{\rm max}=2000$, which represents a conservative choice to avoid potential contamination from small-scale modeling uncertainties, and divide the multipole range into $N_\ell = 20$ logarithmically spaced bins to generate mock observations.
The data used in the work are shown in Figure~\ref{fig:data_ff}, Figure~\ref{fig:data_ww}, and Figure~\ref{fig:data_fw}.
To avoid potential inaccuracies associated with the Limber and flat-sky approximation, the data points with $\ell < 50$ are excluded from the analysis, as illustrated by the shaded regions.
After this selection, the total data vector contains 210 elements, corresponding to a $210\times210$ covariance matrix, 
which is shown in Figure~\ref{fig:cov}.

\begin{deluxetable*}{cccccc}[ht!]
\tablecaption{The fiducial values, priors, and constraint results of the 7 cosmological parameters, 1 baryonic feedback parameter, and 16 systematic parameters in our model derived from the FRB observables, weak lensing measurements, and the joint $3\times2 \,\mathrm{pt}(e,\mathrm{m})$ analysis. The quoted uncertainties correspond to the $1\sigma$ CL, and the relative precision of each parameter is indicated in the bracket.\label{tab:main_result}}
\setlength{\tabcolsep}{4pt}
\tablehead{
\colhead{Parameter} &
\colhead{Fiducial Value} &
\colhead{Prior} &
\colhead{FRBs} &
\colhead{Weak Lensing} &
\colhead{$3\times2 \,\mathrm{pt}(e,\mathrm{m})$}
}
\startdata
\noalign{\vskip 0.1cm}
\multicolumn{6}{c}{\textbf{Cosmological Parameters}}\\
\noalign{\vskip 0.1cm}
\hline
\noalign{\vskip 0.1cm}
$\Omega_{\rm m}$ & 0.3096 & flat $(0.1, 0.9)$ & $0.48^{+0.16}_{-0.23}$ ($63.0\%$) & $0.312\pm0.042$ ($13.6\%$) & $0.326^{+0.031}_{-0.038}$ ($11.1\%$) \\
$\Omega_{\rm b}$ & $0.04897$ & flat $(0.01, 0.1)$ & $0.057\pm0.016$ ($32.7\%$) & $0.0491^{+0.0120}_{-0.0140}$ ($28.5\%$) & $0.0515^{+0.0054}_{-0.0079}$ ($13.6\%$) \\
$h$ & 0.6766 & flat $(0.3, 1.0)$ & $0.64^{+0.15}_{-0.21}$ ($26.6\%$) & $0.706^{+0.083}_{-0.11}$ ($14.2\%$) & $0.681^{+0.053}_{-0.064}$ ($8.6\%$) \\
$n_{\rm s}$ & 0.96 & flat $(0.7, 1.2)$ & $<0.973$ & $0.978^{+0.068}_{-0.046}$ ($5.9\%$) & $0.960^{+0.058}_{-0.037}$ ($5.0\%$) \\
$w$ & -1 & flat $(-2, 0)$ & --- & $-1.02^{+0.22}_{-0.18}$ ($20.0\%$) & $-0.997^{+0.21}_{-0.18}$ ($19.5\%$) \\
$\sum m_{\nu}$ & 0.06 & flat $(0, 2)$ & $<1.2$ & $<0.53$ & $<0.47$ \\
$\sigma_8$ & 0.81 & flat $(0.4, 1.2)$ & $0.86^{+0.31}_{-0.13}$ ($27.2\%$) & $0.784^{+0.040}_{-0.048}$ ($5.4\%$) & $0.782^{+0.037}_{-0.033}$ ($4.3\%$) \\
\noalign{\vskip 0.1cm}
\hline
\noalign{\vskip 0.1cm}
\multicolumn{6}{c}{\textbf{Baryonic Feedback}}\\
\noalign{\vskip 0.1cm}
\hline
\noalign{\vskip 0.1cm}
$\log_{10}T_{\rm AGN}$ & 7.8 & flat $(7.2, 8.4)$ & $7.89^{+0.13}_{-0.08}$ ($1.3\%$) & $7.96^{+0.26}_{-0.23}$ ($3.1\%$) & $7.80\pm0.03$ ($0.4\%$) \\
\noalign{\vskip 0.1cm}
\hline
\noalign{\vskip 0.1cm}
\multicolumn{6}{c}{\textbf{Intrinsic Alignment}}\\
\noalign{\vskip 0.1cm}
\hline
\noalign{\vskip 0.1cm}
$A_{\rm IA}$ & 1 & flat $(-5, 5)$ & --- & $0.97^{+0.16}_{-0.23}$ ($19.5\%$) & $1.03^{+0.15}_{-0.19}$ ($17.0\%$) \\
$\eta_{\rm IA}$ & 0 & flat $(-5, 5)$ & --- & $-0.09^{+0.58}_{-0.73}$ & $0.15\pm0.29$ \\
\noalign{\vskip 0.1cm}
\hline
\noalign{\vskip 0.1cm}
\multicolumn{6}{c}{\textbf{Photo-$z$ Bias}}\\
\noalign{\vskip 0.1cm}
\hline
\noalign{\vskip 0.1cm}
$\Delta z^1$ & 0 & flat $(-0.1, 0.1)$ & --- & $0.004^{+0.032}_{-0.032}$ & $-0.021\pm0.021$ \\
$\Delta z^2$ & 0 & flat $(-0.1, 0.1)$ & --- & $-0.002^{+0.026}_{-0.032}$ & $-0.016\pm0.017$ \\
$\Delta z^3$ & 0 & flat $(-0.1, 0.1)$ & --- & $-0.003^{+0.036}_{-0.046}$ & $-0.009\pm0.024$ \\
$\Delta z^4$ & 0 & flat $(-0.1, 0.1)$ & --- & $-0.005^{+0.045}_{-0.081}$ & $0.004\pm0.042$ \\
$\sigma_z^1/\sigma_{z,{\rm fid}}$ & 1 & flat $(0.5, 1.5)$ & --- & $1.01^{+0.29}_{-0.29}$ & $0.95^{+0.15}_{-0.42}$ \\
$\sigma_z^2/\sigma_{z,{\rm fid}}$ & 1 & flat $(0.5, 1.5)$ & --- & $0.995^{+0.26}_{-0.36}$ & $0.86^{+0.16}_{-0.35}$ \\
$\sigma_z^3/\sigma_{z,{\rm fid}}$ & 1 & flat $(0.5, 1.5)$ & --- & $0.99^{+0.29}_{-0.29}$ & $1.02^{+0.30}_{-0.21}$ \\
$\sigma_z^4/\sigma_{z,{\rm fid}}$ & 1 & flat $(0.5, 1.5)$ & --- & $0.995^{+0.24}_{-0.40}$ & $0.996^{+0.29}_{-0.29}$ \\
\noalign{\vskip 0.1cm}
\hline
\noalign{\vskip 0.1cm}
\multicolumn{6}{c}{\textbf{Shear Calibration}}\\
\noalign{\vskip 0.1cm}
\hline
\noalign{\vskip 0.1cm}
$m_1$ & 0 & flat $(-0.1, 0.1)$ & --- & $0.002^{+0.054}_{-0.046}$ & $0.0059\pm0.0056$ \\
$m_2$ & 0 & flat $(-0.1, 0.1)$ & --- & $0.023^{+0.054}_{-0.035}$ & $0.0055^{+0.0032}_{-0.0036}$ \\
$m_3$ & 0 & flat $(-0.1, 0.1)$ & --- & $0.025^{+0.053}_{-0.033}$ & $0.0029\pm0.0044$ \\
$m_4$ & 0 & flat $(-0.1, 0.1)$ & --- & $0.028^{+0.053}_{-0.033}$ & $0.0019\pm0.0068$ \\
$N^{\gamma \gamma}_{\rm add}$ & $10^{-9}$ & flat $(0.5, 1.5)\times10^{-9}$ & --- & $1.0005^{+0.0029}_{-0.0029}\times10^{-9}$ & $\left(1.0006^{+0.0028}_{-0.0024}\right)\times10^{-9}$ \\
\noalign{\vskip 0.1cm}
\hline
\noalign{\vskip 0.1cm}
\multicolumn{6}{c}{\textbf{DM Noise}}\\
\noalign{\vskip 0.1cm}
\hline
\noalign{\vskip 0.1cm}
$N_{\rm D}$ & 0.54 & flat $(0.1, 0.9)$ & $0.539^{+0.002}_{-0.003}$  & --- & $0.539\pm0.001$  \\
\noalign{\vskip 0.1cm}
\enddata
% \tablecomments{$\sigma_{z,{\rm fid}}=0.05$.}
\end{deluxetable*}

\begin{figure*}[t!]
    \centering    \includegraphics[width=0.9\textwidth]{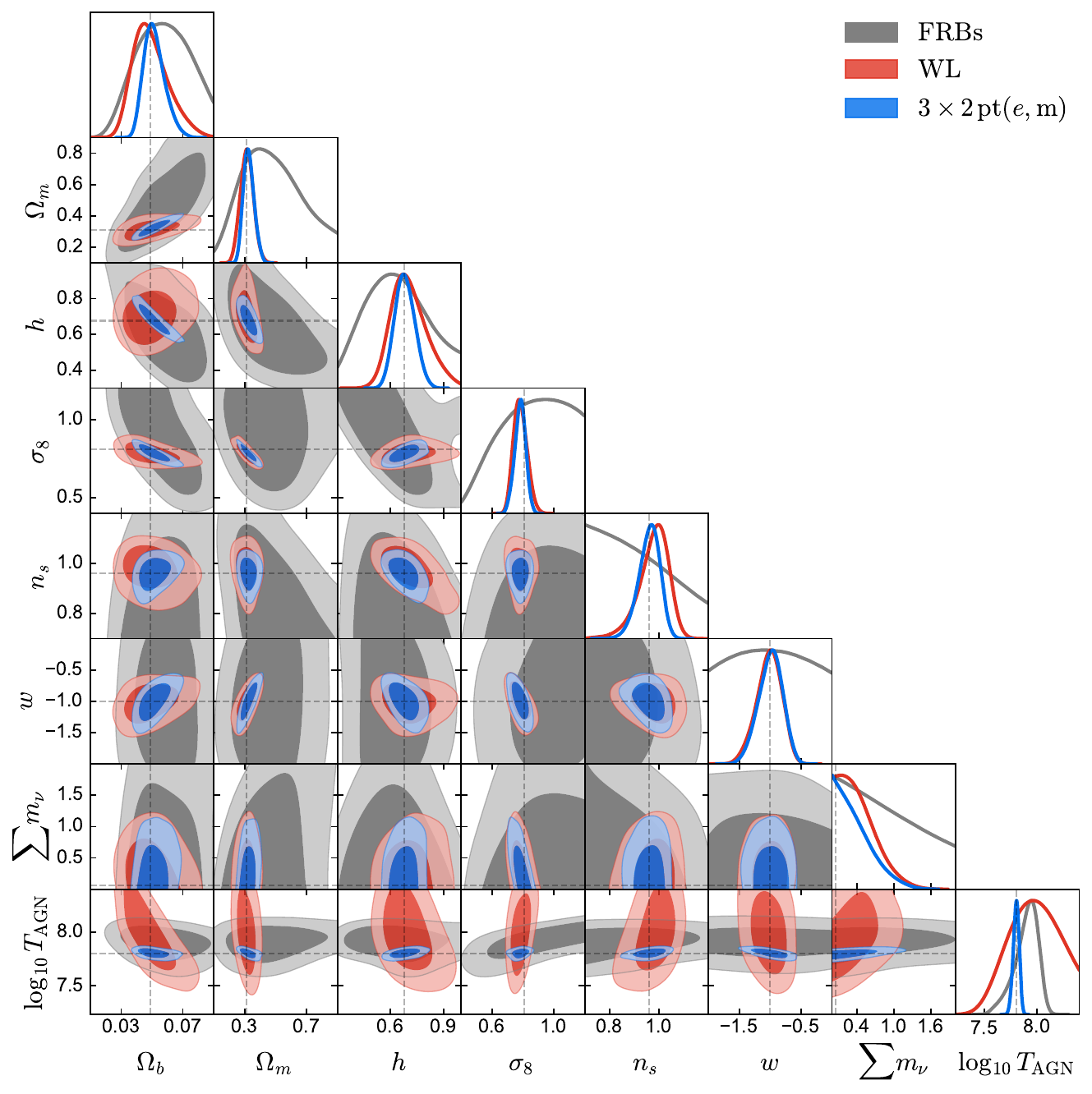}
    \caption{The predicted contour maps (68\% and 95\% CL) and 1D PDFs of the seven cosmological parameters and the baryonic feedback parameter $\log_{10} T_{\rm AGN}$ derived from the FRB observables (gray), CSST weak lensing measurements (red), and the joint $3\times2 \,\mathrm{pt}(e,\mathrm{m})$ analysis (blue). The vertical and horizontal dashed lines indicate the fiducial values of these parameters.}
    \label{fig:main_result}
\end{figure*}

\section{CONSTRAINT AND RESULT}\label{sec:result}
\subsection{Fitting Method}
Using the simulated data from CSST weak lensing surveys and future FRB measurements, 
we perform parameter estimation using the $\chi^2$ method. The $\chi^2$ statistic 
can be written as
\begin{equation}
\chi^2 =
\Big[ \mathbf{D} - \mathbf{M}(\boldsymbol{\theta}) \Big]^{\rm T}
\mathbf{Cov}^{-1}
\Big[ \mathbf{D} - \mathbf{M}(\boldsymbol{\theta}) \Big],
\end{equation}
where $\mathbf{D}$ denotes the simulated data vector, 
$\mathbf{M}(\boldsymbol{\theta})$ represents the corresponding theoretical prediction, 
$\boldsymbol{\theta}$ is the set of model parameters, 
and $\mathbf{Cov}$ is the corresponding covariance matrix. 
The likelihood function is given by $\mathcal{L}
 \propto \exp(-\chi^2/2)$.

Within the Bayesian framework, we use the \texttt{emcee} \citep{2013PASP..125..306F} package to perform 
Markov Chain Monte Carlo (MCMC) sampling to constrain the model parameters. 
We employ 112 independent walkers, each evolving for 30000 steps. 
The first $30\%$ of the samples are discarded as burn-in, and the remaining samples 
from all walkers are combined to construct the posterior distributions of the parameters. 
Our model includes a total of 24 parameters, consisting of 7 cosmological parameters, 
1 baryonic feedback parameter, and 16 systematic parameters. 
The fiducial values and priors of these parameters are listed in 
Table~\ref{tab:main_result}.

\subsection{Constraint results}
In Figure~\ref{fig:main_result}, we present the predicted contour maps and the corresponding marginalized one-dimensional (1D) posterior distribution functions (PDFs) for the cosmological and baryonic feedback parameters, with the 1$\sigma$ (68\%) and 2$\sigma$ (95\%) confidence levels (CL). The best-fit values, $1\sigma$ errors, and corresponding relative precisions of the free parameters derived from the FRBs, weak lensing, and the $3\times2 \,\mathrm{pt}(e,\mathrm{m})$ analyses are summarized in Table~\ref{tab:main_result}.

For the constraint on the baryonic feedback effect, we find that the FRB-only analysis can achieve a relative precision of $1.3\%$ on $\log_{10}T_{\rm AGN}$, significantly outperforming weak lensing alone, which reaches a precision of $3.1\%$. The physical origin of this result lies in the nature of the observables: weak lensing is sensitive to the total matter distribution, which is dominated by dark matter, whereas the FRB DM field directly traces the integrated free electron density along the line of sight. Since the baryon distribution at small scales is much more strongly affected by feedback than the total matter distribution, FRB statistics are uniquely powerful probes of baryonic feedback mechanisms. These mechanisms represent one of the dominant uncertainties in modeling the small-scale matter power spectrum in cosmology \citep{2019OJAp....2E...4C}. 
In the joint $3\times2 \,\mathrm{pt}(e,\mathrm{m})$ analysis, the synergy between FRBs and weak lensing further tightens this constraint to $0.4\%$, representing a substantial improvement over the constraint from weak lensing alone. As the baryonic feedback parameter becomes tightly constrained, the precision of other cosmological parameters can be correspondingly improved.

In our analysis, the effect of massive neutrinos is also considered, which can affect structure formation on small scales and degenerates with the impact of baryonic feedback effect \citep[e.g.][]{2015MNRAS.450.1212H}. In the joint $3\times2 \,\mathrm{pt}(e,\mathrm{m})$ analysis, by precisely constraining the baryonic feedback parameter through FRB DM statistics, we can effectively break the degeneracy and obtain stringent constraint on the sum of neutrino mass $\sum m_{\nu}$. Consequently, the $3\times2 \,\mathrm{pt}(e,\mathrm{m})$ analysis yields a significantly tighter upper limit of $\sum m_{\nu} < 0.47$ eV, demonstrating a notable improvement compared with the weak lensing-only constraint of $\sum m_{\nu} < 0.53$ eV. This result highlights the potential of combining FRBs and weak lensing measurements to constrain the total mass of neutrinos \citep{2025JCAP...12..035B}.

For the other cosmological parameters, weak lensing alone provides strong constraints on $\Omega_{\rm m}$ and $\sigma_8$, but is relatively less sensitive to the baryon density parameter $\Omega_{\rm b}$ and the Hubble parameter $h$. The inclusion of FRB data in the joint $3\times2 \,\mathrm{pt}(e,\mathrm{m})$ analysis effectively breaks parameter degeneracies and improves the overall constraining power. As a result, in the $3\times2 \,\mathrm{pt}(e,\mathrm{m})$ analysis, the constraint precisions on $\Omega_{\rm m}$, $n_s$, $w$, and $\sigma_8$ can reach $11.1\%$, $5.0\%$, $19.5\%$, and $4.3\%$, respectively. In particular, the constraint precisions on $\Omega_{\rm b}$ and $h$ improve significantly, which can reach $13.6\%$ and $8.6\%$, respectively, compared to the results from the weak lensing-only case with the correponding precisions of $28.5\%$ and $14.2\%$.

Besides, we find that the constraints on the intrinsic alignment and some systematic nuisance parameters are also significantly strengthened in the $3\times2 \,\mathrm{pt}(e,\mathrm{m})$ analysis as shown in Figure~\ref{fig:nuisance_params} in Appendix~\ref{app:nuisance_params}. 
% Although FRB observables are not directly related to these parameters, 
The inclusion of FRB information leads to tighter constraints on cosmological and baryonic feedback parameters, which in turn helps break parameter degeneracies within the full parameter space. 
Consequently, several nuisance parameters benefit from the reduced degeneracies and gain tighter bounds, including the intrinsic alignment parameters $A_{\rm IA}$ and $\eta_{\rm IA}$, photo-$z$ calibration parameters $\Delta z_i$, shear calibration parameters $m_i$, and DM field noise parameter $N_{\rm D}$.

\section{SUMMARY AND CONCLUSION}\label{sec:summary}
In this work, we explore the weak lensing measurement from the CSST photometric survey and FRB DM statistics from next-generation radio telescopes (i.e., SKA-Mid or DSA-2000) for probing the baryonic feedback effect. By incorporating various sources of statistical noise and systematic uncertainties, we construct the observed angular power spectra and the corresponding covariance matrix to generate mock data. Within the framework of the $w$CDM cosmological model, we simultaneously consider 7 cosmological parameters, 1 baryonic feedback parameter, and 16 systematic parameters associated with intrinsic alignment, photometric redshift uncertainties, shear calibration and DM noise. Using a MCMC analysis, we forecast the joint constraining power of these observables on these parameters.

We find that CSST weak lensing alone can provide a relatively modest constraint precision of $3.1\%$ on the baryonic feedback parameter $\log_{10}T_{\rm AGN}$, and the result can be dramatically improved to $0.4\%$ by including the FRB DM data from SKA-Mid or DSA-2000 in the joint $3\times2 \,\mathrm{pt}(e,\mathrm{m})$ analysis. By tightly constraining baryonic feedback effect and reducing its degeneracy with other cosmological parameters, the joint analysis enables a more robust inference of the total neutrino mass $\sum m_{\nu}$. As a result, the upper limit is tightened from $\sum m_{\nu}<0.53$ eV for CSST weak lensing alone to $\sum m_{\nu}<0.47$ eV when the future FRB measurement is included.
Besides, the constraints on most of the cosmological and nuisance parameters also can be significantly improved.

Our results highlight the strong potential of FRB DM statistics as a powerful probe of the baryonic feedback effect. With the advent of next-generation radio telescopes and the expected large samples of localized FRBs in the coming decade, FRBs will become increasingly important in the exploration of the Universe. The synergy between future FRB observations and weak lensing surveys will provide a promising tool for accurately studying the baryonic feedback effect, neutrino physics, and other crucial cosmological problems.

\begin{acknowledgments}
S.F. and Y.G. acknowledge the support from the CAS Project for Young Scientists in Basic Research (No. YSBR-092), and National Key R\&D Program of China grant Nos. 2022YFF0503404 and 2020SKA0110402. X.L.C. acknowledges the support of the National Natural Science Foundation of China through grant Nos. 11473044 and 11973047 and the Chinese Academy of Science grants ZDKYYQ20200008, QYZDJ- SSW-SLH017, XDB 23040100, and XDA15020200. This work is also supported by science research grants from the China Manned Space Project with grant Nos. CMS-CSST-2025-A02, CMS-CSST-2021-B01, and CMS-CSST-2021-A01.
\end{acknowledgments}

\appendix
\section{CONSTRAINT RESULTS OF SYSTEMATIC PARAMETERS}\label{app:nuisance_params}
\begin{figure*}[h]
    \centering    \includegraphics[width=1\textwidth]{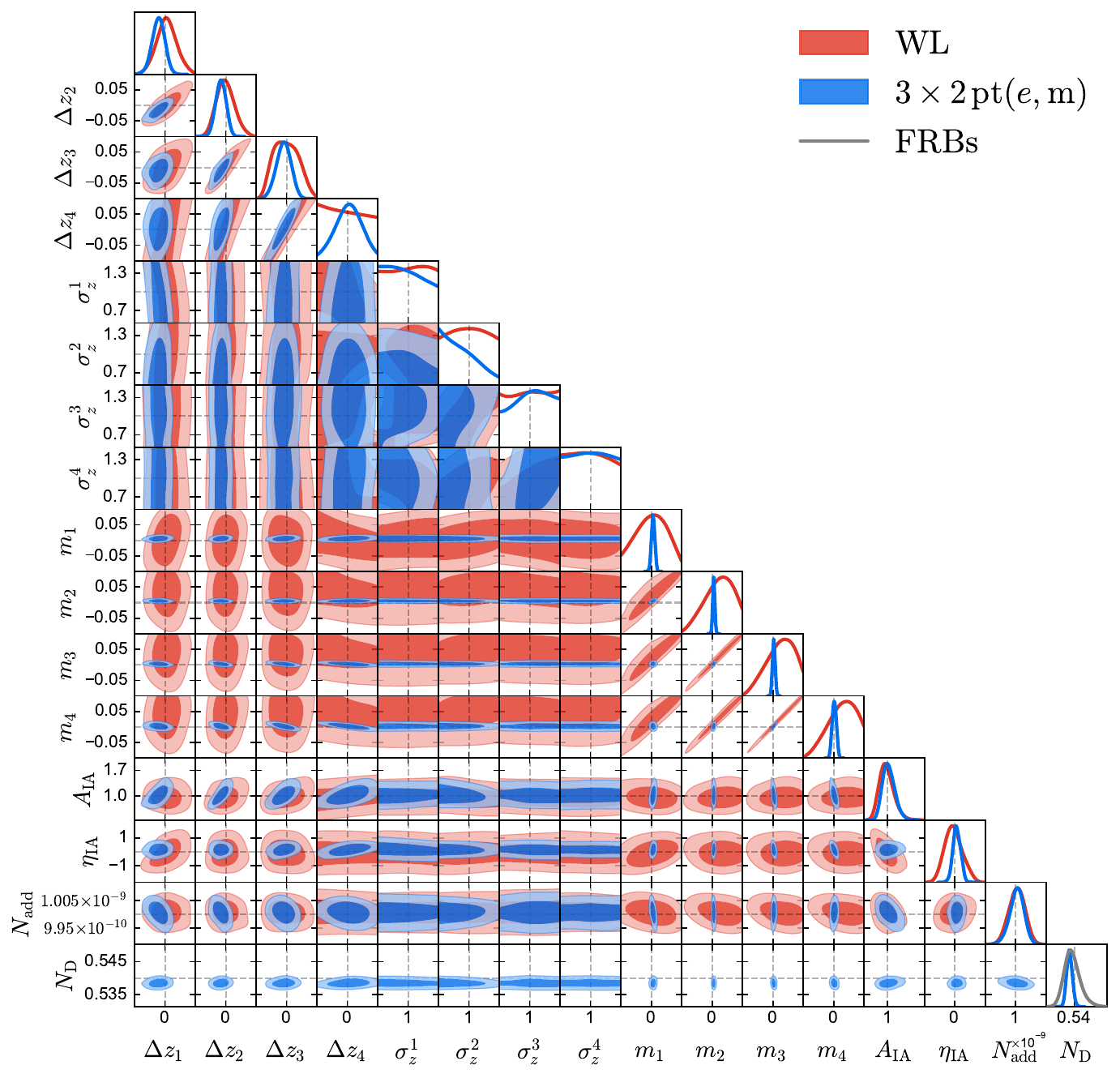}
    \caption{The predicted contour maps (68\% and 95\% CL) and 1D PDFs of the systematic parameters, including the photo-$z$ calibration parameters $\Delta z_i$ and $\sigma_z^i$, shear calibration parameters $m_i$, intrinsic alignment parameters $A_{\rm IA}$ and $\eta_{\rm IA}$, additive error parameter $N_{\rm add}$ and DM field noise parameter $N_{\rm D}$, derived from the FRB observables (gray), CSST weak lensing measurements (red), and $3\times2 \,\mathrm{pt}(e,\mathrm{m})$ analysis (blue). The vertical and horizontal dashed lines indicate the fiducial values of these parameters.}
    \label{fig:nuisance_params}
\end{figure*}

Figure~\ref{fig:nuisance_params} shows the predicted contour maps (68\% and 95\% CL) and 1D PDFs of the systematic parameters forecast from the CSST weak lensing survey and next-generation radio telescope FRB observations, including the photo-$z$ calibration parameters $\Delta z_i$ and $\sigma_z^i$, multiplicative bias parameters $m_i$, intrinsic alignment parameters $A_{\rm IA}$ and $\eta_{\rm IA}$, additive error parameter $N_{\rm add}$ and DM field noise parameter $N_{\rm D}$, derived from the FRB observables (gray), CSST weak lensing measurements (red), and the joint $3\times2 \,\mathrm{pt}(e,\mathrm{m})$ analysis (blue).
\bibliographystyle{aasjournal} 
\bibliography{main}

\begin{thebibliography}{}
\expandafter\ifx\csname natexlab\endcsname\relax\def\natexlab#1{#1}\fi
\providecommand{\url}[1]{\href{#1}{#1}}
\providecommand{\dodoi}[1]{doi:~\href{http://doi.org/#1}{\nolinkurl{#1}}}
\providecommand{\doeprint}[1]{\href{http://ascl.net/#1}{\nolinkurl{http://ascl.net/#1}}}
\providecommand{\doarXiv}[1]{\href{https://arxiv.org/abs/#1}{\nolinkurl{https://arxiv.org/abs/#1}}}

\bibitem[{{Amara} \& {R{\'e}fr{\'e}gier}(2008)}]{2008MNRAS.391..228A}
{Amara}, A., \& {R{\'e}fr{\'e}gier}, A. 2008, \mnras, 391, 228,
  \dodoi{10.1111/j.1365-2966.2008.13880.x}

\bibitem[{{Berghaus} {et~al.}(2025){Berghaus}, {Du}, {Lee}, {Prabhu},
  {Reischke}, {Connor}, \& {Zurek}}]{2025JCAP...12..035B}
{Berghaus}, K.~V., {Du}, Y., {Lee}, V. S.~H., {et~al.} 2025, \jcap, 2025, 035,
  \dodoi{10.1088/1475-7516/2025/12/035}

\bibitem[{{Bochenek} {et~al.}(2021){Bochenek}, {Ravi}, \&
  {Dong}}]{2021ApJ...907L..31B}
{Bochenek}, C.~D., {Ravi}, V., \& {Dong}, D. 2021, \apjl, 907, L31,
  \dodoi{10.3847/2041-8213/abd634}

\bibitem[{{Carilli} \& {Rawlings}(2004)}]{2004NewAR..48..979C}
{Carilli}, C.~L., \& {Rawlings}, S. 2004, \nar, 48, 979,
  \dodoi{10.1016/j.newar.2004.09.001}

\bibitem[{{Cheng} {et~al.}(2026){Cheng}, {Andrew}, {Wang}, \&
  {Masui}}]{2026ApJ...998..252C}
{Cheng}, A.~Q., {Andrew}, S.~E., {Wang}, H., \& {Masui}, K.~W. 2026, \apj, 998,
  252, \dodoi{10.3847/1538-4357/ae369f}

\bibitem[{{Chisari} {et~al.}(2019{\natexlab{a}}){Chisari}, {Mead}, {Joudaki},
  {Ferreira}, {Schneider}, {Mohr}, {Tr{\"o}ster}, {Alonso}, {McCarthy},
  {Martin-Alvarez}, {Devriendt}, {Slyz}, \& {van Daalen}}]{2019OJAp....2E...4C}
{Chisari}, N.~E., {Mead}, A.~J., {Joudaki}, S., {et~al.} 2019{\natexlab{a}},
  The Open Journal of Astrophysics, 2, 4, \dodoi{10.21105/astro.1905.06082}

\bibitem[{{Chisari} {et~al.}(2019{\natexlab{b}}){Chisari}, {Alonso}, {Krause},
  {Leonard}, {Bull}, {Neveu}, {Villarreal}, {Singh}, {McClintock}, {Ellison},
  {Du}, {Zuntz}, {Mead}, {Joudaki}, {Lorenz}, {Tr{\"o}ster}, {Sanchez},
  {Lanusse}, {Ishak}, {Hlozek}, {Blazek}, {Campagne}, {Almoubayyed}, {Eifler},
  {Kirby}, {Kirkby}, {Plaszczynski}, {Slosar}, {Vrastil}, {Wagoner}, \& {LSST
  Dark Energy Science Collaboration}}]{2019ApJS..242....2C}
{Chisari}, N.~E., {Alonso}, D., {Krause}, E., {et~al.} 2019{\natexlab{b}},
  \apjs, 242, 2, \dodoi{10.3847/1538-4365/ab1658}

\bibitem[{{Connor} {et~al.}(2025){Connor}, {Ravi}, {Sharma}, {Ocker}, {Faber},
  {Hallinan}, {Harnach}, {Hellbourg}, {Hobbs}, {Hodge}, {Hodges}, {Kosogorov},
  {Lamb}, {Law}, {Rasmussen}, {Sherman}, {Somalwar}, {Weinreb}, {Woody}, \&
  {Konietzka}}]{2025NatAs...9.1226C}
{Connor}, L., {Ravi}, V., {Sharma}, K., {et~al.} 2025, Nature Astronomy, 9,
  1226, \dodoi{10.1038/s41550-025-02566-y}

\bibitem[{{Cordes} \& {Lazio}(2002)}]{2002astro.ph..7156C}
{Cordes}, J.~M., \& {Lazio}, T.~J.~W. 2002, arXiv e-prints, astro,
  \dodoi{10.48550/arXiv.astro-ph/0207156}

\bibitem[{{CSST Collaboration} {et~al.}(2026){CSST Collaboration}, {Gong},
  {Miao}, {Zhan}, {Li}, {Shangguan}, {Li}, {Liu}, {Chen}, {Yuan}, {Zhou},
  {Liu}, {Yu}, {Ji}, {Qi}, {Liu}, {Dai}, {Wang}, {Zheng}, {Hao}, {Dou}, {Ao},
  {Lin}, {Zhang}, {Wang}, {Sun}, {Li}, {Li}, {Xu}, {Li}, {Li}, {Wu}, {Zhang},
  {Wang}, {Bai}, {Cai}, {Cai}, {Cao}, {Chan}, {Chang}, {Chen}, {Chen}, {Chen},
  {Chen}, {Cui}, {Dong}, {Du}, {Duan}, {Fan}, {Fan}, {Fan}, {Fan}, {Fang},
  {Fu}, {Fu}, {Fu}, {Gao}, {Gu}, {Gu}, {Guo}, {Han}, {Hu}, {Huang}, {Ho},
  {Jiang}, {Jiang}, {Jing}, {Kang}, {Kong}, {Li}, {Li}, {Li}, {Li}, {Li}, {Li},
  {Liao}, {Lin}, {Liu}, {Liu}, {Liu}, {Liu}, {Mao}, {Mao}, {Meng}, {Pang},
  {Peng}, {Peng}, {Shan}, {Shen}, {Shen}, {Shen}, {Shi}, {Shi}, {Tan}, {Tian},
  {Wang}, {Wang}, {Wang}, {Wang}, {Wu}, {Wu}, {Wu}, {Xu}, {Xue}, {Xue}, {Yang},
  {Yang}, {Yao}, {Yuan}, {Yuan}, {Zhang}, {Zhang}, {Zhang}, {Zhang}, {Zhang},
  {Zhao}, {Zhao}, {Zhong}, {Zhong}, {Zhou}, {Zhu}, \&
  {Zu}}]{2026SCPMA..6939501C}
{CSST Collaboration}, {Gong}, Y., {Miao}, H., {et~al.} 2026, Science China
  Physics, Mechanics, and Astronomy, 69, 239501,
  \dodoi{10.1007/s11433-025-2809-0}

\bibitem[{{Debackere} {et~al.}(2020){Debackere}, {Schaye}, \&
  {Hoekstra}}]{2020MNRAS.492.2285D}
{Debackere}, S. N.~B., {Schaye}, J., \& {Hoekstra}, H. 2020, \mnras, 492, 2285,
  \dodoi{10.1093/mnras/stz3446}

\bibitem[{{DES Collaboration} {et~al.}(2026){DES Collaboration}, {Abbott},
  {Aguena}, {Alarcon}, {Alves}, {Amon}, {Anbajagane}, {Andrade-Oliveira},
  {d'Assignies}, {Avila}, {Bacon}, {Beas-Gonzalez}, {Bechtol}, {Becker},
  {Bernstein}, {Blazek}, {Bocquet}, {Brooks}, {Camacho}, {Camacho-Ciurana},
  {Camilleri}, {Campailla}, {Campos}, {Carnero Rosell}, {Carrasco Kind},
  {Carretero}, {Castander}, {Cawthon}, {Chang}, {Choi}, {Coloma-Nadal},
  {Conselice}, {da Costa}, {Costanzi}, {Crocce}, {Davis}, {De Vicente},
  {DePoy}, {DeRose}, {Desai}, {Diehl}, {Doel}, {Doux}, {Drlica-Wagner},
  {Eifler}, {Everett}, {Evrard}, {Fert{\'e}}, {Flaugher}, {Fosalba},
  {Friedrich}, {Frieman}, {Garc{\'\i}a-Bellido}, {Gatti}, {Giannini}, {Giles},
  {Glazebrook}, {Gruen}, {Gruendl}, {Gutierrez}, {Harrison}, {Hartley},
  {Herner}, {Hinton}, {Hollowood}, {Honscheid}, {Huterer}, {Jain}, {James},
  {Jarvis}, {Jeffrey}, {Jeltema}, {Kacprzak}, {Kent}, {Krause}, {Lahav}, {Lee},
  {Legnani}, {Lin}, {Marshall}, {Mau}, {Mena-Fern{\'a}ndez}, {Menanteau},
  {Miquel}, {Mohr}, {Muir}, {Myles}, {Nichol}, {Ogando}, {Palmese}, {Paterno},
  {Percival}, {Petravick}, {Plazas Malag{\'o}n}, {Porredon}, {Prat}, {Preston},
  {Raveri}, {Rodriguez-Monroy}, {Romer}, {Roodman}, {Rykoff}, {Samuroff},
  {S{\'a}nchez}, {Sanchez}, {Sanchez Cid}, {Schutt}, {Sevilla-Noarbe},
  {Sheldon}, {Shin}, {da Silva Pereira}, {Smith}, {Soares-Santos}, {Suchyta},
  {Swanson}, {Tabbutt}, {Tarle}, {Thomas}, {To}, {Troxel}, {Vikram},
  {Vincenzi}, {Weaverdyck}, {Weller}, {Wiseman}, {Yamamoto}, {Yanny}, {Yin}, \&
  {Zuntz}}]{2026arXiv260210065D}
{DES Collaboration}, {Abbott}, T.~M.~C., {Aguena}, M., {et~al.} 2026, arXiv
  e-prints, arXiv:2602.10065, \dodoi{10.48550/arXiv.2602.10065}

\bibitem[{{Euclid Collaboration} {et~al.}(2025){Euclid Collaboration},
  {Mellier}, {Abdurro'uf}, {Acevedo Barroso}, {Ach{\'u}carro}, {Adamek},
  {Adam}, {Addison}, {Aghanim}, {Aguena}, {Ajani}, {Akrami}, {Al-Bahlawan},
  {Alavi}, {Albuquerque}, {Alestas}, {Alguero}, {Allaoui}, {Allen}, {Allevato},
  {Alonso-Tetilla}, {Altieri}, {Alvarez-Candal}, {Alvi}, {Amara}, {Amendola},
  {Amiaux}, {Andika}, {Andreon}, {Andrews}, {Angora}, {Angulo}, {Annibali},
  {Anselmi}, {Anselmi}, {Arcari}, {Archidiacono}, {Aric{\`o}}, {Arnaud},
  {Arnouts}, {Asgari}, {Asorey}, {Atayde}, {Atek}, {Atrio-Barandela}, {Aubert},
  {Aubourg}, {Auphan}, {Auricchio}, {Aussel}, {Aussel}, {Avelino},
  {Avgoustidis}, {Avila}, {Awan}, {Azzollini}, {Baccigalupi}, {Bachelet},
  {Bacon}, {Baes}, {Bagley}, {Bahr-Kalus}, {Balaguera-Antolinez}, {Balbinot},
  {Balcells}, {Baldi}, {Baldry}, {Balestra}, {Ballardini}, {Ballester},
  {Balogh}, {Ba{\~n}ados}, {Barbier}, {Bardelli}, {Baron}, {Barreiro},
  {Barrena}, {Barriere}, {Barros}, {Barthelemy}, {Bartolo}, {Basset},
  {Battaglia}, {Battisti}, {Baugh}, {Baumont}, {Bazzanini}, {Beaulieu},
  {Beckmann}, {Belikov}, {Bel}, {Bellagamba}, {Bella}, {Bellini}, {Benabed},
  {Bender}, {Benevento}, {Bennett}, {Benson}, {Bergamini}, {Bermejo-Climent},
  {Bernardeau}, {Bertacca}, {Berthe}, {Berthier}, {Bethermin}, {Beutler},
  {Bevillon}, {Bhargava}, {Bhatawdekar}, {Bianchi}, {Bisigello}, {Biviano},
  {Blake}, {Blanchard}, {Blazek}, {Blot}, {Bosco}, {Bodendorf}, {Boenke},
  {B{\"o}hringer}, {Boldrini}, {Bolzonella}, {Bonchi}, {Bonici}, {Bonino},
  {Bonino}, {Bonvin}, {Bon}, {Booth}, {Borgani}, {Borlaff}, {Borsato}, {Bose},
  {Botticella}, {Boucaud}, {Bouche}, {Boucher}, {Boutigny}, {Bouvard},
  {Bouwens}, {Bouy}, {Bowler}, {Bozza}, {Bozzo}, {Branchini}, {Brando},
  {Brau-Nogue}, {Brekke}, {Bremer}, {Brescia}, {Breton}, {Brinchmann},
  {Brinckmann}, {Brockley-Blatt}, {Brodwin}, {Brouard}, {Brown}, {Bruton},
  {Bucko}, {Buddelmeijer}, {Buenadicha}, {Buitrago}, {Burger}, {Burigana},
  {Busillo}, {Busonero}, {Cabanac}, {Cabayol-Garcia}, {Cagliari}, {Caillat},
  {Caillat}, {Calabrese}, {Calabro}, {Calderone}, {Calura}, {Camacho Quevedo},
  {Camera}, {Campos}, {Ca{\~n}as-Herrera}, {Candini}, {Cantiello},
  {Capobianco}, {Cappellaro}, {Cappelluti}, {Cappi}, {Caputi}, {Cara},
  {Carbone}, {Cardone}, {Carella}, {Carlberg}, {Carle}, {Carminati}, {Caro},
  {Carrasco}, {Carretero}, {Carrilho}, {Carron Duque}, \&
  {Carry}}]{2025A&A...697A...1E}
{Euclid Collaboration}, {Mellier}, Y., {Abdurro'uf}, {et~al.} 2025, \aap, 697,
  A1, \dodoi{10.1051/0004-6361/202450810}

\bibitem[{{Fedeli}(2014)}]{2014JCAP...04..028F}
{Fedeli}, C. 2014, \jcap, 2014, 028, \dodoi{10.1088/1475-7516/2014/04/028}

\bibitem[{{Fender} {et~al.}(2015){Fender}, {Stewart}, {Macquart}, {Donnarumma},
  {Murphy}, {Deller}, {Paragi}, \& {Chatterjee}}]{2015aska.confE..51F}
{Fender}, R., {Stewart}, A., {Macquart}, J.~P., {et~al.} 2015, in Advancing
  Astrophysics with the Square Kilometre Array (AASKA14), 51,
  \dodoi{10.22323/1.215.0051}

\bibitem[{{Foreman-Mackey} {et~al.}(2013){Foreman-Mackey}, {Hogg}, {Lang}, \&
  {Goodman}}]{2013PASP..125..306F}
{Foreman-Mackey}, D., {Hogg}, D.~W., {Lang}, D., \& {Goodman}, J. 2013, \pasp,
  125, 306, \dodoi{10.1086/670067}

\bibitem[{{Gong} {et~al.}(2019){Gong}, {Liu}, {Cao}, {Chen}, {Fan}, {Li}, {Li},
  {Li}, {Zhang}, \& {Zhan}}]{2019ApJ...883..203G}
{Gong}, Y., {Liu}, X., {Cao}, Y., {et~al.} 2019, \apj, 883, 203,
  \dodoi{10.3847/1538-4357/ab391e}

\bibitem[{{Gong} {et~al.}(2025){Gong}, {Miao}, {Zhou}, {Xiong}, {Song},
  {Jiang}, {Wang}, {Yan}, {Wu}, {Deng}, {Chen}, {Fan}, {Jing}, {Yang}, \&
  {Zhan}}]{2025SCPMA..6880402G}
{Gong}, Y., {Miao}, H., {Zhou}, X., {et~al.} 2025, Science China Physics,
  Mechanics, and Astronomy, 68, 280402, \dodoi{10.1007/s11433-025-2646-2}

\bibitem[{{Hallinan} {et~al.}(2019){Hallinan}, {Ravi}, {Weinreb}, {Kocz},
  {Huang}, {Woody}, {Lamb}, {D'Addario}, {Catha}, {Law}, {Kulkarni}, {Phinney},
  {Eastwood}, {Bouman}, {McLaughlin}, {Ransom}, {Siemens}, {Cordes}, {Lynch},
  {Kaplan}, {Brazier}, {Bhatnagar}, {Myers}, {Walter}, \&
  {Gaensler}}]{2019BAAS...51g.255H}
{Hallinan}, G., {Ravi}, V., {Weinreb}, S., {et~al.} 2019, in Bulletin of the
  American Astronomical Society, Vol.~51, 255,
  \dodoi{10.48550/arXiv.1907.07648}

\bibitem[{{Harnois-D{\'e}raps} {et~al.}(2015){Harnois-D{\'e}raps}, {van
  Waerbeke}, {Viola}, \& {Heymans}}]{2015MNRAS.450.1212H}
{Harnois-D{\'e}raps}, J., {van Waerbeke}, L., {Viola}, M., \& {Heymans}, C.
  2015, \mnras, 450, 1212, \dodoi{10.1093/mnras/stv646}

\bibitem[{{Hildebrandt} {et~al.}(2017){Hildebrandt}, {Viola}, {Heymans},
  {Joudaki}, {Kuijken}, {Blake}, {Erben}, {Joachimi}, {Klaes}, {Miller},
  {Morrison}, {Nakajima}, {Verdoes Kleijn}, {Amon}, {Choi}, {Covone}, {de
  Jong}, {Dvornik}, {Fenech Conti}, {Grado}, {Harnois-D{\'e}raps}, {Herbonnet},
  {Hoekstra}, {K{\"o}hlinger}, {McFarland}, {Mead}, {Merten}, {Napolitano},
  {Peacock}, {Radovich}, {Schneider}, {Simon}, {Valentijn}, {van den Busch},
  {van Uitert}, \& {Van Waerbeke}}]{2017MNRAS.465.1454H}
{Hildebrandt}, H., {Viola}, M., {Heymans}, C., {et~al.} 2017, \mnras, 465,
  1454, \dodoi{10.1093/mnras/stw2805}

\bibitem[{{Hu} {et~al.}(2025){Hu}, {Gong}, {Su}, {Lin}, {Miao}, {Xiong}, \&
  {Chen}}]{2025arXiv251122575H}
{Hu}, D., {Gong}, Y., {Su}, P., {et~al.} 2025, arXiv e-prints,
  arXiv:2511.22575, \dodoi{10.48550/arXiv.2511.22575}

\bibitem[{{Hussaini} {et~al.}(2025){Hussaini}, {Connor}, {Konietzka}, {Ravi},
  {Faber}, {Sharma}, \& {Sherman}}]{2025ApJ...993L..27H}
{Hussaini}, M., {Connor}, L., {Konietzka}, R.~M., {et~al.} 2025, \apjl, 993,
  L27, \dodoi{10.3847/2041-8213/ae0a49}

\bibitem[{{Huterer} {et~al.}(2006){Huterer}, {Takada}, {Bernstein}, \&
  {Jain}}]{2006MNRAS.366..101H}
{Huterer}, D., {Takada}, M., {Bernstein}, G., \& {Jain}, B. 2006, \mnras, 366,
  101, \dodoi{10.1111/j.1365-2966.2005.09782.x}

\bibitem[{{Ivezi{\'c}} {et~al.}(2019){Ivezi{\'c}}, {Kahn}, {Tyson}, {Abel},
  {Acosta}, {Allsman}, {Alonso}, {AlSayyad}, {Anderson}, {Andrew}, {Angel},
  {Angeli}, {Ansari}, {Antilogus}, {Araujo}, {Armstrong}, {Arndt}, {Astier},
  {Aubourg}, {Auza}, {Axelrod}, {Bard}, {Barr}, {Barrau}, {Bartlett}, {Bauer},
  {Bauman}, {Baumont}, {Bechtol}, {Bechtol}, {Becker}, {Becla}, {Beldica},
  {Bellavia}, {Bianco}, {Biswas}, {Blanc}, {Blazek}, {Blandford}, {Bloom},
  {Bogart}, {Bond}, {Booth}, {Borgland}, {Borne}, {Bosch}, {Boutigny},
  {Brackett}, {Bradshaw}, {Brandt}, {Brown}, {Bullock}, {Burchat}, {Burke},
  {Cagnoli}, {Calabrese}, {Callahan}, {Callen}, {Carlin}, {Carlson},
  {Chandrasekharan}, {Charles-Emerson}, {Chesley}, {Cheu}, {Chiang}, {Chiang},
  {Chirino}, {Chow}, {Ciardi}, {Claver}, {Cohen-Tanugi}, {Cockrum}, {Coles},
  {Connolly}, {Cook}, {Cooray}, {Covey}, {Cribbs}, {Cui}, {Cutri}, {Daly},
  {Daniel}, {Daruich}, {Daubard}, {Daues}, {Dawson}, {Delgado}, {Dellapenna},
  {de Peyster}, {de Val-Borro}, {Digel}, {Doherty}, {Dubois},
  {Dubois-Felsmann}, {Durech}, {Economou}, {Eifler}, {Eracleous}, {Emmons},
  {Fausti Neto}, {Ferguson}, {Figueroa}, {Fisher-Levine}, {Focke}, {Foss},
  {Frank}, {Freemon}, {Gangler}, {Gawiser}, {Geary}, {Gee}, {Geha}, {Gessner},
  {Gibson}, {Gilmore}, {Glanzman}, {Glick}, {Goldina}, {Goldstein}, {Goodenow},
  {Graham}, {Gressler}, {Gris}, {Guy}, {Guyonnet}, {Haller}, {Harris},
  {Hascall}, {Haupt}, {Hernandez}, {Herrmann}, {Hileman}, {Hoblitt}, {Hodgson},
  {Hogan}, {Howard}, {Huang}, {Huffer}, {Ingraham}, {Innes}, {Jacoby}, {Jain},
  {Jammes}, {Jee}, {Jenness}, {Jernigan}, {Jevremovi{\'c}}, {Johns}, {Johnson},
  {Johnson}, {Jones}, {Juramy-Gilles}, {Juri{\'c}}, {Kalirai}, {Kallivayalil},
  {Kalmbach}, {Kantor}, {Karst}, {Kasliwal}, {Kelly}, {Kessler}, {Kinnison},
  {Kirkby}, {Knox}, {Kotov}, {Krabbendam}, {Krughoff}, {Kub{\'a}nek},
  {Kuczewski}, {Kulkarni}, {Ku}, {Kurita}, {Lage}, {Lambert}, {Lange},
  {Langton}, {Le Guillou}, {Levine}, {Liang}, {Lim}, {Lintott}, {Long},
  {Lopez}, {Lotz}, {Lupton}, {Lust}, {MacArthur}, {Mahabal}, {Mandelbaum},
  {Markiewicz}, {Marsh}, {Marshall}, {Marshall}, {May}, {McKercher}, {McQueen},
  {Meyers}, {Migliore}, {Miller}, \& {Mills}}]{2019ApJ...873..111I}
{Ivezi{\'c}}, {\v{Z}}., {Kahn}, S.~M., {Tyson}, J.~A., {et~al.} 2019, \apj,
  873, 111, \dodoi{10.3847/1538-4357/ab042c}

\bibitem[{{James} {et~al.}(2022){James}, {Ghosh}, {Prochaska}, {Bannister},
  {Bhandari}, {Day}, {Deller}, {Glowacki}, {Gordon}, {Heintz}, {Marnoch},
  {Ryder}, {Scott}, {Shannon}, \& {Tejos}}]{2022MNRAS.516.4862J}
{James}, C.~W., {Ghosh}, E.~M., {Prochaska}, J.~X., {et~al.} 2022, \mnras, 516,
  4862, \dodoi{10.1093/mnras/stac2524}

\bibitem[{{Joachimi} {et~al.}(2015){Joachimi}, {Cacciato}, {Kitching},
  {Leonard}, {Mandelbaum}, {Sch{\"a}fer}, {Sif{\'o}n}, {Hoekstra}, {Kiessling},
  {Kirk}, \& {Rassat}}]{2015SSRv..193....1J}
{Joachimi}, B., {Cacciato}, M., {Kitching}, T.~D., {et~al.} 2015, \ssr, 193, 1,
  \dodoi{10.1007/s11214-015-0177-4}

\bibitem[{{Joudaki} {et~al.}(2018){Joudaki}, {Blake}, {Johnson}, {Amon},
  {Asgari}, {Choi}, {Erben}, {Glazebrook}, {Harnois-D{\'e}raps}, {Heymans},
  {Hildebrandt}, {Hoekstra}, {Klaes}, {Kuijken}, {Lidman}, {Mead}, {Miller},
  {Parkinson}, {Poole}, {Schneider}, {Viola}, \& {Wolf}}]{2018MNRAS.474.4894J}
{Joudaki}, S., {Blake}, C., {Johnson}, A., {et~al.} 2018, \mnras, 474, 4894,
  \dodoi{10.1093/mnras/stx2820}

\bibitem[{{Kaiser}(1992)}]{1992ApJ...388..272K}
{Kaiser}, N. 1992, \apj, 388, 272, \dodoi{10.1086/171151}

\bibitem[{{Khrykin} {et~al.}(2024){Khrykin}, {Ata}, {Lee}, {Simha}, {Huang},
  {Prochaska}, {Tejos}, {Bannister}, {Cooke}, {Day}, {Deller}, {Glowacki},
  {Gordon}, {James}, {Marnoch}, {Shannon}, {Zhang}, \&
  {Bernales-Cortes}}]{2024ApJ...973..151K}
{Khrykin}, I.~S., {Ata}, M., {Lee}, K.-G., {et~al.} 2024, \apj, 973, 151,
  \dodoi{10.3847/1538-4357/ad6567}

\bibitem[{{Leung} {et~al.}(2025){Leung}, {Borrow}, {Masui}, {Andrew}, {Chen},
  {Schaye}, \& {Schaller}}]{2025arXiv250919514L}
{Leung}, C., {Borrow}, J., {Masui}, K.~W., {et~al.} 2025, arXiv e-prints,
  arXiv:2509.19514, \dodoi{10.48550/arXiv.2509.19514}

\bibitem[{{Lin} {et~al.}(2022){Lin}, {Gong}, {Chen}, {Chan}, {Fan}, \&
  {Zhan}}]{2022MNRAS.515.5743L}
{Lin}, H., {Gong}, Y., {Chen}, X., {et~al.} 2022, \mnras, 515, 5743,
  \dodoi{10.1093/mnras/stac2126}

\bibitem[{{Lin} {et~al.}(2024){Lin}, {Li}, \& {Zou}}]{2024ApJ...969..123L}
{Lin}, H.-N., {Li}, X.-Y., \& {Zou}, R. 2024, \apj, 969, 123,
  \dodoi{10.3847/1538-4357/ad5310}

\bibitem[{{Lorimer} {et~al.}(2007){Lorimer}, {Bailes}, {McLaughlin},
  {Narkevic}, \& {Crawford}}]{2007Sci...318..777L}
{Lorimer}, D.~R., {Bailes}, M., {McLaughlin}, M.~A., {Narkevic}, D.~J., \&
  {Crawford}, F. 2007, Science, 318, 777, \dodoi{10.1126/science.1147532}

\bibitem[{{Lu} {et~al.}(2020){Lu}, {Piro}, \& {Waxman}}]{2020MNRAS.498.1973L}
{Lu}, W., {Piro}, A.~L., \& {Waxman}, E. 2020, \mnras, 498, 1973,
  \dodoi{10.1093/mnras/staa2397}

\bibitem[{{Luo} {et~al.}(2020){Luo}, {Men}, {Lee}, {Wang}, {Lorimer}, \&
  {Zhang}}]{2020MNRAS.494..665L}
{Luo}, R., {Men}, Y., {Lee}, K., {et~al.} 2020, \mnras, 494, 665,
  \dodoi{10.1093/mnras/staa704}

\bibitem[{{Macquart} \& {Ekers}(2018)}]{2018MNRAS.480.4211M}
{Macquart}, J.-P., \& {Ekers}, R. 2018, \mnras, 480, 4211,
  \dodoi{10.1093/mnras/sty2083}

\bibitem[{{Macquart} {et~al.}(2010){Macquart}, {Bailes}, {Bhat}, {Bower},
  {Bunton}, {Chatterjee}, {Colegate}, {Cordes}, {D'Addario}, {Deller},
  {Dodson}, {Fender}, {Haines}, {Halll}, {Harris}, {Hotan}, {Johnston},
  {Jones}, {Keith}, {Koay}, {Lazio}, {Majid}, {Murphy}, {Navarro}, {Phillips},
  {Quinn}, {Preston}, {Stansby}, {Stairs}, {Stappers}, {Staveley-Smith},
  {Tingay}, {Thompson}, {van Straten}, {Wagstaff}, {Warren}, {Wayth}, {Wen}, \&
  {CRAFT Collaboration}}]{2010PASA...27..272M}
{Macquart}, J.-P., {Bailes}, M., {Bhat}, N.~D.~R., {et~al.} 2010, \pasa, 27,
  272, \dodoi{10.1071/AS09082}

\bibitem[{{Macquart} {et~al.}(2020){Macquart}, {Prochaska}, {McQuinn},
  {Bannister}, {Bhandari}, {Day}, {Deller}, {Ekers}, {James}, {Marnoch},
  {Os{\l}owski}, {Phillips}, {Ryder}, {Scott}, {Shannon}, \&
  {Tejos}}]{2020Natur.581..391M}
{Macquart}, J.-P., {Prochaska}, J.~X., {McQuinn}, M., {et~al.} 2020, \nat, 581,
  391, \dodoi{10.1038/s41586-020-2300-2}

\bibitem[{{Margalit} {et~al.}(2019){Margalit}, {Berger}, \&
  {Metzger}}]{2019ApJ...886..110M}
{Margalit}, B., {Berger}, E., \& {Metzger}, B.~D. 2019, \apj, 886, 110,
  \dodoi{10.3847/1538-4357/ab4c31}

\bibitem[{{Martizzi} {et~al.}(2013){Martizzi}, {Teyssier}, \&
  {Moore}}]{2013MNRAS.432.1947M}
{Martizzi}, D., {Teyssier}, R., \& {Moore}, B. 2013, \mnras, 432, 1947,
  \dodoi{10.1093/mnras/stt297}

\bibitem[{{McCarthy} {et~al.}(2017){McCarthy}, {Schaye}, {Bird}, \& {Le
  Brun}}]{2017MNRAS.465.2936M}
{McCarthy}, I.~G., {Schaye}, J., {Bird}, S., \& {Le Brun}, A. M.~C. 2017,
  \mnras, 465, 2936, \dodoi{10.1093/mnras/stw2792}

\bibitem[{{McQuinn}(2014)}]{2014ApJ...780L..33M}
{McQuinn}, M. 2014, \apjl, 780, L33, \dodoi{10.1088/2041-8205/780/2/L33}

\bibitem[{{Mead} {et~al.}(2020){Mead}, {Tr{\"o}ster}, {Heymans}, {Van
  Waerbeke}, \& {McCarthy}}]{2020A&A...641A.130M}
{Mead}, A.~J., {Tr{\"o}ster}, T., {Heymans}, C., {Van Waerbeke}, L., \&
  {McCarthy}, I.~G. 2020, \aap, 641, A130, \dodoi{10.1051/0004-6361/202038308}

\bibitem[{{Metzger} {et~al.}(2017){Metzger}, {Berger}, \&
  {Margalit}}]{2017ApJ...841...14M}
{Metzger}, B.~D., {Berger}, E., \& {Margalit}, B. 2017, \apj, 841, 14,
  \dodoi{10.3847/1538-4357/aa633d}

\bibitem[{{Miyatake} {et~al.}(2023){Miyatake}, {Sugiyama}, {Takada},
  {Nishimichi}, {Li}, {Shirasaki}, {More}, {Kobayashi}, {Nishizawa}, {Rau},
  {Zhang}, {Takahashi}, {Dalal}, {Mandelbaum}, {Strauss}, {Hamana}, {Oguri},
  {Osato}, {Luo}, {Kannawadi}, {Hsieh}, {Armstrong}, {Bosch}, {Komiyama},
  {Lupton}, {Lust}, {MacArthur}, {Miyazaki}, {Murayama}, {Okura}, {Price},
  {Sunayama}, {Tait}, {Tanaka}, \& {Wang}}]{2023PhRvD.108l3517M}
{Miyatake}, H., {Sugiyama}, S., {Takada}, M., {et~al.} 2023, \prd, 108, 123517,
  \dodoi{10.1103/PhysRevD.108.123517}

\bibitem[{{Navarro} {et~al.}(1997){Navarro}, {Frenk}, \&
  {White}}]{1997ApJ...490..493N}
{Navarro}, J.~F., {Frenk}, C.~S., \& {White}, S. D.~M. 1997, \apj, 490, 493,
  \dodoi{10.1086/304888}

\bibitem[{{Ocker} \& {Cordes}(2026)}]{2026arXiv260211838O}
{Ocker}, S.~K., \& {Cordes}, J.~M. 2026, arXiv e-prints, arXiv:2602.11838,
  \dodoi{10.48550/arXiv.2602.11838}

\bibitem[{{Pakmor} {et~al.}(2023){Pakmor}, {Springel}, {Coles}, {Guillet},
  {Pfrommer}, {Bose}, {Barrera}, {Delgado}, {Ferlito}, {Frenk}, {Hadzhiyska},
  {Hern{\'a}ndez-Aguayo}, {Hernquist}, {Kannan}, \&
  {White}}]{2023MNRAS.524.2539P}
{Pakmor}, R., {Springel}, V., {Coles}, J.~P., {et~al.} 2023, \mnras, 524, 2539,
  \dodoi{10.1093/mnras/stac3620}

\bibitem[{{Rafiei-Ravandi} {et~al.}(2020){Rafiei-Ravandi}, {Smith}, \&
  {Masui}}]{2020PhRvD.102b3528R}
{Rafiei-Ravandi}, M., {Smith}, K.~M., \& {Masui}, K.~W. 2020, \prd, 102,
  023528, \dodoi{10.1103/PhysRevD.102.023528}

\bibitem[{{Rafiei-Ravandi} {et~al.}(2021){Rafiei-Ravandi}, {Smith}, {Li},
  {Masui}, {Josephy}, {Dobbs}, {Lang}, {Bhardwaj}, {Patel}, {Bandura},
  {Berger}, {Boyle}, {Brar}, {Breitman}, {Cassanelli}, {Chawla}, {Adam Dong},
  {Fonseca}, {Gaensler}, {Giri}, {Good}, {Halpern}, {Kaczmarek}, {Kaspi},
  {Leung}, {Lin}, {Mena-Parra}, {Meyers}, {Michilli}, {M{\"u}nchmeyer}, {Ng},
  {Petroff}, {Pleunis}, {Rahman}, {Sanghavi}, {Scholz}, {Shin}, {Stairs},
  {Tendulkar}, {Vanderlinde}, \& {Zwaniga}}]{2021ApJ...922...42R}
{Rafiei-Ravandi}, M., {Smith}, K.~M., {Li}, D., {et~al.} 2021, \apj, 922, 42,
  \dodoi{10.3847/1538-4357/ac1dab}

\bibitem[{{Reischke} \& {Hagstotz}(2025)}]{2025arXiv250717742R}
{Reischke}, R., \& {Hagstotz}, S. 2025, arXiv e-prints, arXiv:2507.17742,
  \dodoi{10.48550/arXiv.2507.17742}

\bibitem[{{Reischke} {et~al.}(2023){Reischke}, {Neumann}, {Bertmann},
  {Hagstotz}, \& {Hildebrandt}}]{2023arXiv230909766R}
{Reischke}, R., {Neumann}, D., {Bertmann}, K.~A., {Hagstotz}, S., \&
  {Hildebrandt}, H. 2023, arXiv e-prints, arXiv:2309.09766,
  \dodoi{10.48550/arXiv.2309.09766}

\bibitem[{{Schaye} {et~al.}(2023){Schaye}, {Kugel}, {Schaller}, {Helly},
  {Braspenning}, {Elbers}, {McCarthy}, {van Daalen}, {Vandenbroucke}, {Frenk},
  {Kwan}, {Salcido}, {Bah{\'e}}, {Borrow}, {Chaikin}, {Hahn}, {Hu{\v{s}}ko},
  {Jenkins}, {Lacey}, \& {Nobels}}]{2023MNRAS.526.4978S}
{Schaye}, J., {Kugel}, R., {Schaller}, M., {et~al.} 2023, \mnras, 526, 4978,
  \dodoi{10.1093/mnras/stad2419}

\bibitem[{{Schneider} \& {Teyssier}(2015)}]{2015JCAP...12..049S}
{Schneider}, A., \& {Teyssier}, R. 2015, \jcap, 2015, 049,
  \dodoi{10.1088/1475-7516/2015/12/049}

\bibitem[{{Sharma} {et~al.}(2026{\natexlab{a}}){Sharma}, {Krause}, {Ravi},
  {Reischke}, {Connor}, {R.~S.}, \& {Anbajagane}}]{2026ApJ...998..109S}
{Sharma}, K., {Krause}, E., {Ravi}, V., {et~al.} 2026{\natexlab{a}}, \apj, 998,
  109, \dodoi{10.3847/1538-4357/ae2ff9}

\bibitem[{{Sharma} {et~al.}(2025){Sharma}, {Krause}, {Ravi}, {Reischke},
  {R.~S.}, \& {Connor}}]{2025ApJ...989...81S}
---. 2025, \apj, 989, 81, \dodoi{10.3847/1538-4357/adeca4}

\bibitem[{{Sharma} {et~al.}(2026{\natexlab{b}}){Sharma}, {Krause}, {Ravi},
  {Anbajagane}, {Connor}, {Kimmy Wu}, {Ferraro}, {Grandis}, {Alonso}, {Chiang},
  {Law}, {Pranjal R.}, {McCarty}, \& {Pandey}}]{2026arXiv260422105S}
---. 2026{\natexlab{b}}, arXiv e-prints, arXiv:2604.22105,
  \dodoi{10.48550/arXiv.2604.22105}

\bibitem[{{Shirasaki} {et~al.}(2017){Shirasaki}, {Kashiyama}, \&
  {Yoshida}}]{2017PhRvD..95h3012S}
{Shirasaki}, M., {Kashiyama}, K., \& {Yoshida}, N. 2017, \prd, 95, 083012,
  \dodoi{10.1103/PhysRevD.95.083012}

\bibitem[{{Shirasaki} {et~al.}(2022){Shirasaki}, {Takahashi}, {Osato}, \&
  {Ioka}}]{2022MNRAS.512.1730S}
{Shirasaki}, M., {Takahashi}, R., {Osato}, K., \& {Ioka}, K. 2022, \mnras, 512,
  1730, \dodoi{10.1093/mnras/stac490}

\bibitem[{{Shirasaki} {et~al.}(2026){Shirasaki}, {Takahashi}, {Osato}, \&
  {Ioka}}]{2026arXiv260121336S}
---. 2026, arXiv e-prints, arXiv:2601.21336, \dodoi{10.48550/arXiv.2601.21336}

\bibitem[{{SKAO Science Team}(2015)}]{ska_lev0}
{SKAO Science Team}. 2015, {SKA1 Level 0 Science Requirements}, Tech. Rep.
  SKA-TEL-SKO-0000007-Rev02, SKAO

\bibitem[{{Su} {et~al.}(2026){Su}, {Gong}, {Xiong}, {Hu}, {Lin}, {Deng}, \&
  {Chen}}]{2026ApJ..1000..143S}
{Su}, P., {Gong}, Y., {Xiong}, Q., {et~al.} 2026, \apj, 1000, 143,
  \dodoi{10.3847/1538-4357/ae4a1e}

\bibitem[{{Sun} {et~al.}(2015){Sun}, {Zhang}, \& {Li}}]{2015ApJ...812...33S}
{Sun}, H., {Zhang}, B., \& {Li}, Z. 2015, \apj, 812, 33,
  \dodoi{10.1088/0004-637X/812/1/33}

\bibitem[{{Takahashi} {et~al.}(2025){Takahashi}, {Ioka}, {Shirasaki}, \&
  {Osato}}]{2025arXiv251102155T}
{Takahashi}, R., {Ioka}, K., {Shirasaki}, M., \& {Osato}, K. 2025, arXiv
  e-prints, arXiv:2511.02155, \dodoi{10.48550/arXiv.2511.02155}

\bibitem[{{Theis} {et~al.}(2024){Theis}, {Hagstotz}, {Reischke}, \&
  {Weller}}]{2024arXiv240308611T}
{Theis}, A., {Hagstotz}, S., {Reischke}, R., \& {Weller}, J. 2024, arXiv
  e-prints, arXiv:2403.08611, \dodoi{10.48550/arXiv.2403.08611}

\bibitem[{{Tr{\"o}ster} {et~al.}(2022){Tr{\"o}ster}, {Mead}, {Heymans}, {Yan},
  {Alonso}, {Asgari}, {Bilicki}, {Dvornik}, {Hildebrandt}, {Joachimi},
  {Kannawadi}, {Kuijken}, {Schneider}, {Shan}, {van Waerbeke}, \&
  {Wright}}]{2022A&A...660A..27T}
{Tr{\"o}ster}, T., {Mead}, A.~J., {Heymans}, C., {et~al.} 2022, \aap, 660, A27,
  \dodoi{10.1051/0004-6361/202142197}

\bibitem[{{van Daalen} {et~al.}(2011){van Daalen}, {Schaye}, {Booth}, \& {Dalla
  Vecchia}}]{2011MNRAS.415.3649V}
{van Daalen}, M.~P., {Schaye}, J., {Booth}, C.~M., \& {Dalla Vecchia}, C. 2011,
  \mnras, 415, 3649, \dodoi{10.1111/j.1365-2966.2011.18981.x}

\bibitem[{{Van Waerbeke} {et~al.}(2000){Van Waerbeke}, {Mellier}, {Erben},
  {Cuillandre}, {Bernardeau}, {Maoli}, {Bertin}, {McCracken}, {Le F{\`e}vre},
  {Fort}, {Dantel-Fort}, {Jain}, \& {Schneider}}]{2000A&A...358...30V}
{Van Waerbeke}, L., {Mellier}, Y., {Erben}, T., {et~al.} 2000, \aap, 358, 30,
  \dodoi{10.48550/arXiv.astro-ph/0002500}

\bibitem[{{Wanderman} \& {Piran}(2015)}]{2015MNRAS.448.3026W}
{Wanderman}, D., \& {Piran}, T. 2015, \mnras, 448, 3026,
  \dodoi{10.1093/mnras/stv123}

\bibitem[{{Wang} {et~al.}(2025){Wang}, {Masui}, {Andrew}, {Fonseca},
  {Gaensler}, {Joseph}, {Kaspi}, {Kharel}, {Lanman}, {Leung}, {Mas-Ribas},
  {Mena-Parra}, {Nimmo}, {Pearlman}, {Pen}, {Prochaska}, {Raikman}, {Shin},
  {Siegel}, {Smith}, \& {Stairs}}]{2025arXiv250608932W}
{Wang}, H., {Masui}, K., {Andrew}, S., {et~al.} 2025, arXiv e-prints,
  arXiv:2506.08932, \dodoi{10.48550/arXiv.2506.08932}

\bibitem[{{Wayland} {et~al.}(2026){Wayland}, {Alonso}, \&
  {Reischke}}]{2026MNRAS.tmp..515W}
{Wayland}, A., {Alonso}, D., \& {Reischke}, R. 2026, \mnras,
  \dodoi{10.1093/mnras/stag557}

\bibitem[{{Wei} \& {Gao}(2024)}]{2024ApJ...975..184W}
{Wei}, J.-J., \& {Gao}, C.-Y. 2024, \apj, 975, 184,
  \dodoi{10.3847/1538-4357/ad82e4}

\bibitem[{{Wright} {et~al.}(2025){Wright}, {St{\"o}lzner}, {Asgari}, {Bilicki},
  {Giblin}, {Heymans}, {Hildebrandt}, {Hoekstra}, {Joachimi}, {Kuijken}, {Li},
  {Reischke}, {von Wietersheim-Kramsta}, {Yoon}, {Burger}, {Chisari}, {de
  Jong}, {Dvornik}, {Georgiou}, {Harnois-D{\'e}raps}, {Jalan}, {William},
  {Joudaki}, {Lesci}, {Linke}, {Loureiro}, {Mahony}, {Maturi}, {Miller},
  {Moscardini}, {Napolitano}, {Porth}, {Radovich}, {Schneider}, {Tr{\"o}ster},
  {Valentijn}, {Wittje}, {Yan}, \& {Zhang}}]{2025A&A...703A.158W}
{Wright}, A.~H., {St{\"o}lzner}, B., {Asgari}, M., {et~al.} 2025, \aap, 703,
  A158, \dodoi{10.1051/0004-6361/202554908}

\bibitem[{{Xiong} {et~al.}(2026){Xiong}, {Gong}, {Yan}, {Deng}, {Lin}, {Zhou},
  {Chen}, {Guo}, {Li}, {Liu}, \& {Pei}}]{2026ApJ...998..320X}
{Xiong}, Q., {Gong}, Y., {Yan}, J., {et~al.} 2026, \apj, 998, 320,
  \dodoi{10.3847/1538-4357/ae3a8d}

\bibitem[{{Yan} {et~al.}(2026){Yan}, {Gong}, {Xiong}, {Chen}, {Guo}, {Li},
  {Liu}, \& {Pei}}]{2026ApJ...997..357Y}
{Yan}, J.-h., {Gong}, Y., {Xiong}, Q., {et~al.} 2026, \apj, 997, 357,
  \dodoi{10.3847/1538-4357/ae2d09}

\bibitem[{{Yao} {et~al.}(2017){Yao}, {Manchester}, \&
  {Wang}}]{2017ApJ...835...29Y}
{Yao}, J.~M., {Manchester}, R.~N., \& {Wang}, N. 2017, \apj, 835, 29,
  \dodoi{10.3847/1538-4357/835/1/29}

\bibitem[{{Y{\"u}ksel} {et~al.}(2008){Y{\"u}ksel}, {Kistler}, {Beacom}, \&
  {Hopkins}}]{2008ApJ...683L...5Y}
{Y{\"u}ksel}, H., {Kistler}, M.~D., {Beacom}, J.~F., \& {Hopkins}, A.~M. 2008,
  \apjl, 683, L5, \dodoi{10.1086/591449}

\bibitem[{{Zhan}(2011)}]{2011SSPMA..41.1441Z}
{Zhan}, H. 2011, Scientia Sinica Physica, Mechanica \& Astronomica, 41, 1441,
  \dodoi{10.1360/132011-961}

\bibitem[{Zhan(2021)}]{Zhan2021TheWM}
Zhan, H. 2021, Chinese Science Bulletin.
\newblock \url{https://api.semanticscholar.org/CorpusID:234805827}

\bibitem[{{Zhang}(2023)}]{2023RvMP...95c5005Z}
{Zhang}, B. 2023, Reviews of Modern Physics, 95, 035005,
  \dodoi{10.1103/RevModPhys.95.035005}

\bibitem[{{Zhang} {et~al.}(2023){Zhang}, {Zhao}, {Li}, {Zhang}, {Li}, \&
  {Zhang}}]{2023SCPMA..6620412Z}
{Zhang}, J.-G., {Zhao}, Z.-W., {Li}, Y., {et~al.} 2023, Science China Physics,
  Mechanics, and Astronomy, 66, 120412, \dodoi{10.1007/s11433-023-2212-9}

\bibitem[{{Zhang} {et~al.}(2021){Zhang}, {Zhang}, {Li}, \&
  {Lorimer}}]{2021MNRAS.501..157Z}
{Zhang}, R.~C., {Zhang}, B., {Li}, Y., \& {Lorimer}, D.~R. 2021, \mnras, 501,
  157, \dodoi{10.1093/mnras/staa3537}

\bibitem[{{Zhang} \& {Zhang}(2026)}]{2026arXiv260704792Z}
{Zhang}, Z.-L., \& {Zhang}, B. 2026, arXiv e-prints, arXiv:2607.04792,
  \dodoi{10.48550/arXiv.2607.04792}

\end{thebibliography}
\end{document}